\documentstyle[12pt,aaspp,psfig]{article}
\slugcomment{\shortstack[r]{Accepted by MNRAS}}
\begin{document}
\newcommand\Msun {M_{\odot}\ }
\newcommand\Lsun {L_{\odot}\ }

\title{Using the Ca~II Triplet to Trace 
Abundance Variations in Individual Red Giant Branch stars
in Three Nearby Galaxies\footnotemark[1]}
\vskip0.5cm

\author{{\bf Eline Tolstoy}\footnotemark[2]$^,$\footnotemark[3]}
\affil{UK GEMINI Support Group, Nuclear and Astrophysics Laboratory, 
University of Oxford, Keble Road, Oxford OX1 3RH, UK}

\author{{\bf Michael J. Irwin\footnotemark[4]}}
\affil{Institute of Astronomy, University of Cambridge, 
Madingley Road, Cambridge, CB3 0HA, UK}

\author{{\bf Andrew A. Cole \footnotemark[5]}}
\affil{Department of Astronomy, 532A LGRT, University of Massachusetts, 
Amherst, MA~01003, USA}

\author{{\bf L. Pasquini \& R. Gilmozzi \footnotemark[6]}} 
\affil{European Southern Observatory, Karl-Schwarzschild str. 2,
85748 Garching bei M\"{u}nchen, Germany}

\author{{\bf J.S. Gallagher \footnotemark[7]}}
\affil{Department of Astronomy, University of Wisconsin-Madison, 
475 North Charter Street, Madison, WI 53706, USA}

\footnotetext[1]{Based on observations collected at the European Southern
Observatory, proposal numbers 63.N-0603 and 61.A-0275}
\footnotetext[2]{Present address: Kapteyn Institute, 
Postbus 800, 9700AV Groningen, the Netherlands}
\footnotetext[3]{email: etolstoy@astro.ox.ac.uk}
\footnotetext[4]{email: mike@ast.cam.ac.uk}
\footnotetext[5]{email: cole@condor.astro.umass.edu}
\footnotetext[6]{email: lpasquin@eso.org, rgilmozz@eso.org}
\footnotetext[7]{email: jsg@astro.wisc.edu}

\newpage

\begin{abstract}

Spectroscopic abundance determinations for stars spanning a Hubble
time in age are necessary in order to unambiguously determine the
evolutionary histories of galaxies.  Using FORS1 in Multi-Object
Spectroscopy mode on ANTU (UT1) at the ESO-VLT on Paranal we obtained
near infrared spectra from which we measured the equivalent widths of
the two strongest Ca~II triplet lines to determine metal abundances
for a sample of Red Giant Branch stars, selected from ESO-NTT optical
(I, V$-$I) photometry of three nearby, Local Group, galaxies: the
Sculptor Dwarf Spheroidal, the Fornax Dwarf Spheroidal and the Dwarf
Irregular NGC~6822.  The summed equivalent width of the two strongest
lines in the Ca~II triplet absorption line feature, centered at
8500~\AA , can be readily converted into an [Fe/H] abundance using the
previously established calibrations by Armandroff \& Da~Costa (1991)
and Rutledge, Hesser \& Stetson (1997).  We measured metallicities for
37 stars in Sculptor, 32 stars in Fornax, and 23 stars in NGC 6822,
yielding more precise estimates of the metallicity distribution
functions for these galaxies than it is possible to obtain
photometrically.  In the case of NGC 6822, this is the first direct
measurement of the abundances of the intermediate-age and old stellar
populations.  We find metallicity spreads in each galaxy which are
broadly consistent with the photometric width of the Red Giant Branch,
although the abundances of individual stars do not always appear to
correspond to their colour.  This is almost certainly predominantly
due to a highly variable star formation rate with time in these
galaxies, which results in a non-uniform, non-globular-cluster-like,
evolution of the Ca/Fe ratio.

\end{abstract}


\newpage

\section{Introduction}

It is impossible to uniquely determine the star formation history of a
galaxy on the basis of photometry of individual Red Giant Branch (RGB)
stars alone. This is because the uncertainty in the metallicity (heavy
element abundance) of a star translates into an uncertainty in age.
This is the age-metallicity degeneracy which plagues the accurate
analysis of CMDs ({\it e.g.}, Searle, Wilkinson \& Bagnuolo 1980;
Tolstoy 1998, and references therein; Cole {\it et al.} 1999).  It is
necessary to directly measure the abundance of stars of different ages
to understand how metallicity has changed with time. This has not been
easily possible for galaxies beyond the Magellanic Clouds until the
advent of 8m class telescopes, even at the intermediate resolutions
required for the most basic of metallicity indicators, the Ca~II
triplet. Looking at isolated systems beyond our rather complex
immediate neighbourhood means that we can avoid the difficult
interpretations required to understand the properties of our Galaxy
({\it e.g.}, Edvardsson {\it et al.} 1993) and the Magellanic Clouds
({\it e.g.}, Olszewski {\it et al.} 1991; Da~Costa \& Hatzidimitriou
1998; Cole, Smecker-Hane \& Gallagher 2000) and find systems that are
simpler and thus (hopefully) easier to interpret. We will most
definitely extend our knowledge to a larger variety of star formation
histories.

One of the major uncertainties in galaxy evolution remains our
detailed understanding of how the abundance of heavy elements, or
metallicity, in an interstellar medium varies with time, and thus in
different generations of stars. In galaxies which are still forming
stars today, the end-point of the metallicity evolution can be
measured using H~II region emission lines ({\it e.g.}, Matteucci \&
Tosi 1985; Pagel \& Tautvai\v{s}ien\.e 1998), or super-giant stars
({\it e.g.}, Venn {\it et al.} 2000), but deducing how the metal
abundances built up to their present levels requires additional
information.  We want to be able to measure the evolution of
metallicity directly and consistently through time, by looking at the
properties of stars of different ages.  Using a Colour-Magnitude
Diagram (CMD) of individual stars in a galaxy it is possible to select
RGB stars with ages in the range 1$-$10~Gyr and by determining their
individual metallicities we can monitor the metallicity evolution of
the whole galaxy over this time frame using the same index.  If we
then add this independent metallicity information about individual
stars to the CMD analysis, then we can better disentangle the effects
of age and metallicity and determine a more accurate star formation
history over the age range 1$-$10~Gyr ({\it e.g.}, Cole {\it et al.}
2000; Hughes \& Wallerstein 2000; Brown \& Wallerstein 1993).

The Ca~II triplet lines, at 8498, 8542 and 8662 \AA \, are
conveniently among the strongest features in the near-infrared spectra
of most late-type stars, and only moderate spectral resolution is
required to accurately measure their strengths.  The use of the Ca~II
triplet as a metallicity indicator has a long and checkered history,
but after some initial uncertainty, it was shown, in the integrated
light of a sample of globular clusters, that the Ca~II triplet summed
equivalent width is strongly affected by metallicity ({\it e.g.},
Armandroff \& Zinn 1988). Subsequent studies backed-up this result
from measurements of individual globular cluster RGB stars ({\it
e.g.}, Olszewski {\it et al.} 1991).  An empirical method of ranking
globular clusters according to metallicity was first detailed by
Armandroff \& Da~Costa (1991, hereafter AD91), and their basic
approach is what has been most generally adopted and refined since
then.  Rutledge {\it et al.} (1997a) have presented the most extensive
catalogue of Ca~II triplet measurements, which Rutledge, Hesser \&
Stetson (1997b, hereafter R97b) have used to calibrate the Ca~II
triplet, [Ca/H] to the [Fe/H] scale determined from detailed high
resolution spectroscopy by Zinn \& West (1984) and Carretta \& Gratton
(1997). R97b have shown that the Ca~II triplet method is accurate and
linear between [Fe/H]$= -2.2$ and $-0.6$, which is the range of
metallicities we might expect for the galaxies in our sample.

Thus, measuring the summed strengths of the two strongest Ca~II
triplet lines in the spectra of individual RGB stars and converting
them into a measure of stellar iron abundance is an empirically proven
``quick and dirty'' alternative to high resolution detailed direct
abundance determinations of numerous elements ({\it e.g.}, AD91;
Suntzeff et al. 1993; R97b; Da Costa \& Hatzidimitriou 1998; Cole {\it
et al.} 2000), if we can assume that globular clusters and galaxy
field stars will have similar abundance patterns 
(Smecker-Hane \& McWilliam 1999; Shetrone, C\^{o}t\'{e} \& Sargent 2001).

With FORS1 in multi-object spectroscopy (MOS) mode we can efficiently
build up a large sample of Ca~II triplet measurements of individual
stars in nearby galaxies.  Using the Ca~II triplet lines we can also
determine the radial velocity of each star and thus assess the
likelihood of the membership in the galaxy, and obtain a rough
determination of the velocity dispersion of the observed stars within
these galaxies.

We chose to observe three nearby galaxies which are known to have
complex, long-lasting star formation histories. The resulting large
range in age of stars on the RGB makes it very complicated to directly
interpret its properties in terms of age or metallicity.  The three
nearby galaxies chosen are: Sculptor dSph, Fornax dSph and NGC~6822 dI
(see Table~1). These galaxies have accurate CMDs, and quite complex,
but relatively well determined, star formation histories ({\it e.g.},
Monkiewicz {\it et al.} 1999; Buonanno {\it et al.} 1999; Gallart et
al. 1996a).
%Table 1 contains the name of the galaxy in column 1; the galactic
%coordinates in columns 2, 3; the V magnitude of the Horizontal Branch
%in column 4; the distance modulus in column 5; the reddening in column
%6; the absolute magnitude of the galaxy in column 7; the radial
%velocity, and the sigma on this value in columns 8, 9; the type of
%galaxy in column 10 and the references from which these data were
%taken in column 11.
Stars in four relatively nearby star clusters with well determined
metal abundances and previous Ca~II triplet observations were also
observed (see Table~2) to calibrate the variation of the Ca~II triplet
lines as a function of metallicity,
and thus tie our results onto the metallicity scale of R97b. 
%Table~2 contains the name of the star cluster in column 1; the
%galactic coordinates in columns 2, 3; the Iron abundance in column 4;
%the reddening in column 5; the V magnitude of the Horizontal Branch in
%column 6; the distance modulus in column 6 and the radial velocity of
%the cluster in column 7.

\newpage

\section{Observations}

The observations were obtained in visitor-mode, with UT1/FORS1 with
MOS instrument set-up, between August 17th and 20th 1999 (see
Table~3).
%Table~3 lists the date of the observation in column 1; the time they
%began in UT in column 2; the Object name in column 3; the exposure
%time in column 4; the airmass in column 5; and the seeing as measured
%by the DIMM in column 6; and any comments in column 7.  
At our resolution, the FORS1 MOS field of view is 6.8 arcmin long and
$\sim$2 arcmin wide to cover the full wavelength range (7000 - 9000
\AA \,). It is covered by 19 mechanical slit-jaws which can be moved
around the field horizontally for a given orientation on the sky.  The
slit jaws come from either side of the field to meet at the determined
slit width. The length of each slit is fixed, and in the configuration we
used each slit is about 20 arcsec long (projected on the sky).

Through-out our observations we used a slit width of 1 arcsec and
the GRIS-600I+15 grism along with the OG590 order-sorting filter, to
cover the Ca~II triplet wavelength region with as high resolution as
possible.  With this setting the pixel sampling is close to 1\AA \,
per pixel and the resolution $\approx$2-3\AA \, over the wavelength
range 7050$-$9150\AA\,.  This is the maximum resolution that can be
obtained with FORS1 without resorting to a narrower slit. Although
this is a wavelength range at which the FORS1 CCD (Tektronix) has
reduced sensitivity it is where the RGB stars we were aiming to
detect are brightest. The Ca~II triplet is also a useful unblended
feature to accurately measure radial velocities ({\it e.g.,} Hargreaves
{\it et al.} 1994) and there are abundant narrow sky lines in this
region for wavelength calibration and/or spectrograph flexure
monitoring.  The spectrum in the region of the Ca~II triplet is also
very flat and relatively free of other lines, permitting unambiguous
continuum level determination.

The meteorological conditions during this run were always photometric
(see Tolstoy {\it et al.} 2000), and the seeing although very good
typically varied by quite a lot during each of the nights (roughly
between 0.3 and 0.9 arcsec on the seeing monitor)
%We tried to concentrate on this
%programme during periods of (relatively) poor seeing. The excellent
%seeing conditions were used to image nearby galaxies (Tolstoy et
%al. 2000; Tolstoy {\it et al.} in prep.).  
These data were taken while Paranal was still under construction,
before the re-coating of the primary mirror, and were therefore
30$-$40\% below the optimal sensitivity.

The target RGB stars were selected from earlier NTT imaging in V and I
filters of these galaxies (Tolstoy {\it et al.} in prep).  We selected
stars from three different fields in Sculptor (see
Figure~\ref{sclcont}) and one each in Fornax (Figure~\ref{fnxcont})
and in NGC6822 (Figure~\ref{n6822cont}). The selection criteria
covered a range in magnitude and we attempted to get as large as
possible a spread in colour across the RGB (See Figures~\ref{sclcmd},
\ref{fnxcmd}, \ref{n6822cmd}). We also tried to avoid including
asymptotic giant branch stars by avoiding the tip region of the RGB,
although it has recently been shown (Cole {\it et al.} 2000) that this
does not have an important impact on the results.  Our selection
within these criteria was then driven by how best to align the FORS1
MOS field with the available candidates.

Astrometry of the selected targets was determined through the NTT
images in combination with the FORS1 pre-imaging. Through-slit images
were used to check the crucial centering of objects in their slits.
%We made FORS1 pre-images of each of our spectroscopic fields to
%optimise the accuracy with which we can place the MOS slits precisely
%on the desired target over the whole field of view.  Although a
%reasonably accurate WCS could be added to the NTT images, it was found
%to be more reliable to over plot the NTT positions on FORS1 pre-images
%and then determine the MOS field set-up using the FORS image and the
%FORS Instrumental Mask Simulator (FIMS) tool. This then automatically
%took care of any distortions in the FORS1 field.  Also, as part of
%standard observing practice on Paranal a through-slit image is always
%taken before commencing spectroscopic exposures.  
We have been able to use these images to further check for small
offsets from centre for the final positions of all the objects over
the field. This is crucial to accurately determine radial velocities
of all the stars across the MOS field. A shift of one pixel of the
object from the centre of a 5 pixel wide slit translates into $\sim$35
km/s error in a radial velocity determination.  As we show later in
the paper, we were able to correct for the small offsets that are
inevitable with this kind of complex multi-slit setup. We also used
the through-slit image to correct for uncertain photometry for some of
the globular cluster stars.

Membership of each individual star observed in each of the galaxies
was assumed based upon their photometric positions in a
Colour-Magnitude Diagram. This was then verified from radial velocity
determinations based upon the resulting spectra. Happily, there was
little evidence for contamination by foreground (or background!)
objects in either Sculptor or Fornax nor, within the large
uncertainties, NGC~6822.

\section{Data Reduction and Analysis}

As part of standard observing practice on Paranal bias frames are
taken every day, as are internal flat fields and wavelength
calibration arc spectra, which are taken through the MOS set-ups 
used at night.

We reduced all our data in IRAF\footnote{IRAF is distributed by the
National Optical Astronomy Observatories, which are operated by the
Association of Universities for Research in Astronomy, Inc., under
cooperative agreement with the National Science Foundation.}, using
standard routines from the CCDRED and APALL packages. We de-biased our
frames making use of the overscan region, and flat fielded them from
the day-time calibration flat field frames.

Working in the wavelength range 7050--9150\AA \, at a resolution of
1\AA /pixel, the night sky lines are plentiful and well distributed
over the wavelength range and were used to directly map the wavelength
distortion and accurately calibrate the spectra.  The adopted
reference wavelengths were taken from the on-line Keck LRIS skyline
plots, which were in turn based on a compilation by Osterbrock \&
Martel (1992). This was found to be more accurate than using the
day-time arc spectra.  The sky is estimated from regions as
symmetrically disposed as possible either side of the target in the
(typically) 20arcsec long slits.  The resolution of the spectrograph
is 2-3 pixels in the spectral direction and 0.2 arcsec/pixel in the
spatial direction therefore sky subtraction presents no problem.

%\newpage

\subsection{Determining Membership $-$ Radial Velocities}

We determined the radial velocities of all the observed stars using
the techniques previously described in Tolstoy \& Irwin (2000).  We
created a radial velocity template from Ruprecht 106 - star 1614,
which has a previous accurate radial velocity measured by Da~Costa et
al. (1992) to be $-$54 km/s. This was used as the zero-point
comparison for all the radial velocities quoted in this paper,
cross-correlating this template with all the other spectra using
FXCOR. We further checked the velocity system using the four stars in
Pal~12 with previously measured radial velocities (see Table~4).

One the largest errors in determining radial velocities with this MOS
set-up comes from errors in centering the images in the slits.  These
systematic radial velocity errors can be corrected by determining the
offsets of the star position in a through-slit image, and comparing
this to a cross-correlation with the position of the telluric A band
in their respective spectra to confirm the displacement and correct
the wavelength scale appropriately.
%Therefore for each set-up the through-slit image 
%was inspected to look for evidence of offset of the images in
%the slit. We also cross-correlated the A-band region (around 7600 \AA\
%) of the spectra with the template to look for evidence that this
%atmospheric line shifted around due to offsets in position of the star
%in the slit.  This a very broad line, and it is thus not always very
%accurate method to determine small offsets on sub-pixel scales, but it
%was a useful confirmation of what was seen in the through-slit
%image. This also allowed us to check any absolute offset in radial
%velocity due to FORS1 flexure between the Rup~106 star (taken at
%airmass 1.7), and the typical observations of the target galaxies at
%air-masses between 1 and 1.1. This effect is going to be very small
%(see Tolstoy \& Irwin 2000), but non-negligible for accurate radial
%velocities with the resolution of FORS.  The corrections determined in
%this way were then applied to the measured radial velocities.

As shown in Figures~\ref{sclhis}, \ref{fnxhis} \& \ref{n6822his} for
each galaxy the histogram of radial velocity tightens up significantly
about the (known) radial velocity of the galaxy after the correction
has been applied. The expected stellar velocity dispersion within
globular clusters and dwarf spheroidals is known to be about $\pm$~10
km/s ({\it e.g.}, Harris 1996; Mateo 1998), but within larger dI
galaxies this is not so well known, and is also much less meaningful,
as dIs typically rotate. In HI gas NGC~6822 has a velocity difference
of 100 km/s from one side to the other (de Blok \& Walter 2000;
Brandenburg \& Skillman 1998), and the velocity dispersion we see in
Figure~\ref{n6822his} is a combination of the effects of differential
rotation over the field and random motion.

Radial velocities were derived for all the spectra mainly for use in
determining membership probabilities.  Rather than using 
radial velocity standards we calibrated our velocity system using the
known radial velocities of stars within globular clusters we also
observed. This also has the added advantage of
providing a better template match for cross-correlation.  The majority
of the RGB-selected stars in each MOS field ought to be members of the
galaxies because of the careful photometric selection.  In combination
with the radial velocity information this generally leads to
unambiguous membership assignment.  For Fornax and Sculptor previous
studies at high resolution have shown the dispersion to be around 10
km/s (Armandroff \& Da~Costa 1986; Queloz, Dubath \& Pasquini 1995;
Mateo {\it et al.} 1991).  Based upon the dispersion we measure from
our spectra, we determine our criteria for membership such that within
3 sigma of the central velocity is a ``definite'' member, and in the
range 3$-$5 sigma is a ``maybe'', and those (very few) outside this
range are highly unlikely to be members of the system.

In the case of Fornax and NGC~6822 there is likely to be a degree of
foreground Galactic star contamination which cannot be eliminated by
radial velocity information. The velocity of Fornax is very low, and
thus it is frequently hard to distinguish members of Fornax dSph from
stars in our Galaxy (see Mateo {\it et al.} 1991), but careful
colour selection helps to minimise this contamination.
Fornax is at high galactic latitude, so this effect is in any case
likely to be small, but NGC~6822 has
both a low velocity and a low Galactic latitude.  It is liable to have
significant contamination which we will be unable to detect
directly, see also Gallart {\it et al.} (1996a).  This can clearly be seen
in our NGC~6822 CMD (in Figure~\ref{n6822cmd}), compared to Fornax
(Figure~\ref{fnxcmd}) and Sculptor (Figure~\ref{sclcmd}), there are 
obviously field stars over the entire area of the
NGC~6822 RGB. This is seen as a scatter of points covering the
magnitude and colour range plotted in Figure~\ref{n6822cmd}.  NGC~6822
also suffers from significant reddening without much information upon
the differential effects. Nonetheless, the distribution of velocities
of the RGB stars selected in NGC~6822 is still peaked about the
expected value of the galaxy (from Richter, Tammann \& Huchtmeier
1987), and so this gives us confidence that the contamination is
likely to be a small fraction of the total. 

\subsection{Determining the Abundance $-$ Equivalent Widths}

We measured equivalent widths of the Ca~II triplet lines in IRAF by
fitting a Gaussian profile to each line and determining the continuum
consistently on either side of the line.  The weakest Ca~II line at
8498 \AA \, was not used in any analysis.  It is often of very low
signal-to-noise, especially at low metallicity, and is more affected
by sky lines than the other two.  The error estimates on the Gaussian
fits as provided by the fitting routine are given in Tables (4$-$7).
The errors in the equivalent width due to inaccurate continuum
placement in the Ca~II triplet region are much less than the random
errors.  We also checked for a possible offset caused by the Gaussian
fitting function by comparing these results with a simple total
integration over the lines and found negligible ($<$~1\%) systematic
differences.  We present example spectra with different line widths
(and thus metallicity) in Figure~\ref{spec}.

The dependence of the Ca~II line widths on surface gravity, for
metallicities below [Ca/H]$= -0.3$ on the RGB above the Horizontal
Branch, has been shown to be a simple linear relationship between
W$^\prime$, which we define as the summed width of
$\lambda\lambda$8542 and 8662, and the absolute magnitude, M$_V$, of
the star which can be parametrized (V$-$V$_{HB}$) the difference in
magnitude between the observed RGB star and the level of the
Horizontal Branch ({\it e.g.}, AD91, R97b).  So this dependence can be
easily removed, and we are left with a straight forward dependence of
W$^\prime$ on metallicity.  In common with Cole {\it et al.} (2000),
we re-derived the R97b relation, which requires the minimum of
assumptions and input data:

$W^\prime = \Sigma W(Ca) + 0.64(\pm 0.02)(V - V_{HB})$

\noindent{where} $\Sigma W(Ca) = W_{8542} + W_{8662}$, which is very
similar to the linear regime of the AD91 calibration. We also adopt
the abundance scale of Carretta \& Gratton (1997), which was shown by
R97b to scale linearly with $W'$ as:

$[Fe/H]_{CG97} = -2.66 + 0.42W^\prime$

\noindent{We} rederived this relation using observations of individual
stars in globular clusters of known metallicity. We can then apply this
abundance scale to our observations of individual stars in 
nearby galaxies.

There are a number of uncertainties in putting the stars observed in
galaxies on the same basis as the measurement of globular clusters, as
extensively discussed in R97a, and Cole {\it et al.} (2000). The most
important is the [Ca/Fe] ratio, which is likely to vary depending upon
the star formation history of the galaxy, and the ratio of old to
young stellar populations.  The [Fe/H] abundance at a given age should
be related to the total amount of past star-formation (until that
age), but [Ca/Fe] at a given age is a function of the ratio of star
formation during past 100 Myr prior to that age, and the total star
formation until that age.  This means that care must be taken in
converting between Ca and Fe abundances, and a Galactic-halo-like
calibration is suitable for those cases when the total star-formation
is dominated by the very earliest times ({\it e.g.}, Draco and Ursa
Minor, Shetrone {\it et al.} 2001), but a Galactic-disc-like
calibration is better for a more constant star-formation rate ({\it
e.g.}, Smecker-Hane \& Mc~William~1999 for the Sagittarius dSph). When
a galaxy is dominated by discrete episodes of star-formation,
especially in recent times it becomes difficult to uniquely determine
a unique metallicity scale from a single element abundance ({\it
e.g.}, Pagel \& Tautvai\v{s}ien\.e 1998, for the SMC; and Cole {\it et
al.} 2000). 

There is also the problem of choosing the correct value of V$_{HB}$,
which varies systematically with age in a galaxy with extended periods
of star-formation (Cole 1998).  In stellar populations much younger
than the calibrating globular clusters, the corresponding V$_{HB}$
will likely be brighter than predicted for given metallicity ({\it
e.g.}, Da Costa \& Hatzidimitriou 1998).  This effect is exacerbated
in a system with young stars and variable reddening, such as NGC~6822.
For field stars, with an a priori unknown age, we can't do anything
about this fact, which could add a bias of order 0.1 dex to the
metallicity of an individual star.  However, this does not preclude a
detailed analysis of the results, provided care is taken in the
interpretation (Cole {\it et al.} 2000).

\section{Results}

\subsection{Calibration Globular Clusters}

To calibrate our observations on to a metallicity scale, as
demonstrated in the comprehensive studies of globular clusters by AD91
and R97b we observed individual stars in four globular clusters: M~15,
Rup~106, Pal~12 and 47~Tuc, which cover the metallicity range we are
interested in (see Table~2), and which have previously been studied
and put into a global scheme by AD91 and R97b.  There appear to be
inherent differences in the measurement of equivalent widths between
previous studies, probably due to differences in line fitting
techniques and telescope and instrument properties (see R97a,b for
detailed discussion), and so we redetermine the scale using our own
observations.

The results of our observations of stars in globular clusters are
listed in Table~4, which shows: the cluster and star number in column
1; the V magnitude of the star in column 2; the B-V colour in column
3; our measured summed equivalent width of the two strongest Ca~II
triplet lines, $\Sigma $W$_{ob}$ = W$_{8542}$ + W$_{8662}$ in column
4; the uncorrected observed radial velocity, v$_r$(ob) in column 5;
the correction to the observed radial velocity due to the offset of
the star in the slit in column 6; the corrected velocity, after this
offset has been applied, V$_r$(c) in column 7; a previous measurement
of the radial velocity of the star if one exists in column 8; the
references for the colour, magnitude and any previous observations for
each star are listed in column 9.

In Figure~\ref{globs} the summed Ca~II equivalent widths are plotted
as a function of V$-$V$_{HB}$ for the stars in our calibration
clusters. For each cluster the line of constant metallicity determined
for each cluster is plotted, using the R97b calibration described in
\S3.2, as a dashed line.

Broadly speaking our results fit consistently onto the previous
abundance scale determined by R97b, even though we define W$^\prime$
differently (R97a use a weighted sum of all three triplet lines).  The
mean metallicity of Rup~106 is found to be somewhat higher ([Fe/H]$=
-1.4$) than that determined by Da~Costa {\it et al.} 1992 ([Fe/H]$=
-1.7$), but it is more consistent with the value measured by Brown,
Wallerstein \& Zucker (1997) from high resolution spectroscopy.  In
the case of M~15, we find it to be slightly more metal rich ([Fe/H]$=
-2.15$) than expected from high resolution spectroscopy, where Sneden
{\it et al.} (1997) find values around [Fe/H]$= -2.3$.  Sneden {\it et
al.}  noted that M~15 has an unusual high [Ca/Fe] ratio, which could
explain this discrepancy.  Pal~12 appears to have a slightly higher
mean metallicity ([Fe/H]$= -0.8$) than high resolution spectroscopy
found ([Fe/H]$= -1.0$), from Brown {\it et al.} (1997), but consistent
with previous Ca~II triplet observations (R97a). Our data also support
the metallicity spread noted by Brown {\it et al.} due to variations
in the [Ca/Fe] ratio within this cluster.

It is known that the [Ca/Fe] ratio is not really that constant over
the globular cluster population (Carney 1996).  Therefore there is an
implicit assumption in the use of the (AD91, R97b) calibrations that
the objects of interest have experienced a similar evolution of the
[Ca/Fe] to [Fe/H] ratio as the globular clusters.  Figure~8 of Cole
{\it et al.} (2000) shows how this assumption may break down for the
LMC, because of its complex star-formation history.  Until the Ca~II
triplet equivalent width can be calibrated to a scale which ties it
directly to [Ca/H], this remains the ultimate limitation on the
accuracy of the method.

\subsection{Sculptor Dwarf Spheroidal}

\subsubsection{Background}

The Sculptor dSph galaxy was discovered by Shapley in 1938 (Shapley
1938), and Baade \& Hubble (1939) noted its similarity to a globular
cluster, except for size and distance. Sculptor was found to contain a
very rich population of RR Lyr variable stars (Thackeray 1950),
clearly indicating that its stellar population contains a globular
cluster age component.  Various studies of evolved stars have shown
Sculptor to contain a small number (8) of intermediate age carbon
stars ({\it e.g.}, Frogel {\it et al.} 1982; Azzopardi {\it et al.}
1986), well known indicators of metal poor intermediate age stellar
populations ({\it e.g.}, Aaronson {\it et al.} 1984) and possibly also
Anomalous Cepheids (Norris \& Bessel 1978).  The presence of old and
intermediate age stellar populations clearly demonstrates that
Sculptor is predominantly old, but there are signs that it has been
forming some stars until at least 7$-$8~Gyr ago. Carignan {\it et al.}
(1998) found evidence for small amounts HI gas in and around Sculptor.

The first attempt to piece together an accurate star-formation history
determined from a CMD reaching down to globular cluster age main
sequence turnoffs was made by Da~Costa~(1984).  He determined that the
bulk of the stellar population of Sculptor is likely to be 2$-$3~Gyr
younger than a typical globular cluster, such as M~92.  He also noted
that the intrinsic width of the RGB was probably caused by a 0.5~dex
spread in abundance (from [Fe/H] = $-$2.1 to $-$1.6), and a population
of ``blue-stragglers'' (in globular cluster terminology), which could
be interpreted as main sequence turn-off stars as young as 5~Gyr old,
or they could be the results of stellar mergers.  Kaluzny et
al. (1995) made a careful study of the central region of Sculptor and
found no main sequence stars (m$_V < 21$), ruling out any star
formation over the last 2~Gyrs.

Using WFPC2, Monkiewicz {\it et al.} (1999) have made the deepest CMD of
Sculptor to date, although they cover a tiny fraction of the entire
galaxy, they detected stars several magnitudes below the oldest
possible main sequence turnoffs.  Their accurate photometry in this
region allowed them to conclude that the mean age of Sculptor is
similar to that of a globular cluster, but that there was probably a
spread in age during this epoch of at least 4~Gyr.

%Sculptor contains an unusually large number of red horizontal branch
%stars for such an intrinsically metal poor system, and there have been
%various attempts to explain this ``second parameter'' effect.
%Da~Costa~(1984) suggested that age might create this effect, given
%that Sculptor is likely to be younger than an average globular
%cluster, with a large spread in its epoch of formation.  Recently
%Majewski {\it et al.} (1999) suggested that there is a bimodality in
%Sculptor's metallicity distribution, and the two different populations
%have different radial distributions, and this effectively gives
%Sculptor an {\it internal} second-parameter problem.  Hurley-Keller,
%Mateo \& Grebel (1999) similarly look for an internal mechanism for
%the second parameter problem, but speculate a radial gradient in the
%binary-star population to be the cause.

\subsubsection{Ca~II Triplet Results}

In Table~5 we summarise the results for the sample of stars we
observed in Sculptor in the 3 fields across the galaxy, distributed as
shown in Figure~\ref{sclcont}. The nomenclature for the stars in
Table~5 is that stars with c1 or c2 before a number are taken from the
central field (a distinction is made between the two susi2 chips, {\it
i.e.}, c1 (East) and c2 (West). The NE field is o1 and the SW one is
o2.  The two outer fields were chosen to overlap with the Carignan et
al. (1998) detections of HI.  The position of the selected stars in a
CMD are shown in Figure~\ref{sclcmd}.

The Ca~II triplet results are plotted on the equivalent width versus
the V magnitude difference with the Horizontal Branch, calibrated to
the R97b results in Figure~\ref{sclres}.  Only stars which are
considered to be members of Sculptor from their radial velocity, and
lie on the RGB are fully included in Figure~\ref{sclcmd}. 
%We have
%plotted the location (with crosses) of where the rejected stars would
%lie (see Table~5).  
We have also excluded a number of stars because of
their positions in the CMD. One star has a spectrum which looks like
an M-star, one lies in the Horizontal Branch region of the CMD, and
four are too far to the blue side of the RGB, and are unlikely to be
RGB stars. Since the Ca~II triplet has only been shown to be valid for
RGB stars it is sensible to exclude these stars from our
analysis. They are also plotted as crosses in Figure~\ref{sclres}.

It is clear that the Red Giants across Sculptor contain a significant
spread in metallicity, see Figure~\ref{sclfehist}.  The highest
metallicity appears to be around [Fe/H]$= - 1.3$, and there are a few
objects around [Fe/H]$ = -$2.1, but the majority are clustered between
[Fe/H]$ = -$2.0 and $-$1.3.  
The mean metallicity of the stars we observed in 
Sculptor is [Fe/H]$= - 1.5 \pm 0.3$.
There is no evidence for any spatial
effect in the distribution of stars of different metallicity, and each
of the observed fields (c1, c2, o1, and o2) contain stars which fall
over the entire range of metallicity found for Sculptor. The histogram
plot shows a fairly sharp cut off in the upper metallicity boundary
for Sculptor, with a shallow tail extending to low metallicity.

In general terms the Ca~II triplet results give us a larger
metallicity spread than we would expect from the breadth of the RGB
observed in CMDs, as would be expected, if the metal-poor stars are
older than the metal-rich stars.  Most of the stars seem to cluster
around [Fe/H]$\sim -$1.5, which is slightly more metal rich than the
RGB suggested.  This serves as a useful example of the inherent
uncertainties caused by the age-metallicity degeneracy in determining
metallicities from the RGB.

Another important point of note is that our results show that the
colour of an individual star in Sculptor on the RGB is {\bf not} a
very good indicator of metallicity, not even relative metallicity
across the RGB. There are blue stars from the metal rich side of the
metallicity distribution and red stars from the metal poor side. There
is however a slight general trend in the population of stars of
different metallicity to be more red with higher metallicities, but
there is a lot of dispersion.  This also means that we often find
isochrones of the metallicity of the star we observe do not fit the
V$-$I colour of the star for any age. This reconfirms the fairly well
known fact that we do not understand stellar evolution on the RGB very
well.

\subsubsection{An Evolutionary Scenario}

One possible global star formation history scenario for Sculptor which
takes into account all the information that we now have on the 
stellar population, is that Sculptor has under-gone two distinct, 
possibly contiguous, phases of star-formation during its life time.

The initial phase, was probably the most intense. If the original gas
was pristine then the enrichment was extremely rapid, because the mean
metallicity of the stars from this epoch seems to be [Fe/H]$\sim
-1.7$.  The duration of this phase of active star formation was from
15$-$11~Gyr ago, effectively an extended period of star-formation
beginning about the age of Galactic globular cluster formation, and
continuing for around 4~Gyr. This assumption comes from the deep CMD
analysis of main sequence turnoffs by Monkiewicz et al. 1999, and also
the analysis of the horizontal branch morphology being caused by an
age spread at early time ({\it e.g.}, Da~Costa 1984; Majewski {\it et
al.} 1999).

The second phase of star formation in Sculptor is required to explain
the population of intermediate age evolved stars such as carbon
stars, AGB stars and Anomalous Cepheids. The average metallicity of
these stars is [Fe/H]$\sim -1.5$. It is hard to give an older age
limit to these stars, but theory dictates that they are typically
younger than 9~Gyr, and the limits on the brightest main sequence
objects makes the younger age limit for star formation 4$-$5~Gyr ago.

Whether these two phases are actually part of a continuous, though
declining star formation rate between 15 and 5 Gyr ago, or
representative of two distinct epochs, or perhaps ``bursts'' it is
hard to say for sure. It is only clear that star formation younger
than about 11~Gyr ago has to be much less intense than the period
between 11-15~Gyr ago because of the sparsely populated main sequence
at these ages, and relatively few intermediate age stars.

Putting together all the pieces of information we have about past star
formation in Sculptor, one possible star formation history is plotted
in the upper panel of Figure~\ref{sclsfh} as a dashed line.  
The dashed line in the lower panel describes the most simple chemical
evolution model (as first described by Searle \& Sargent 1972),
assuming the star formation history in the upper panel. We derived
the assumed yield (the rate at which stars are producing Fe) by 
taking as fixed the high metallicity end
point of the star formation history at the most recent time.
On the lower plot we then over-plot where
the RGB stars would lie, if we determine ages for the stars for which
we measured the metallicity using isochrones (Bertelli {\it et al.} 1994).
Our results are consistent with a trend of increasing metallicity with
time since formation, but there is a large scatter at all ages.
There appears to a broad agreement between the trend seen in the
data points and in the model.

\subsection{Fornax Dwarf Spheroidal}

\subsubsection{Background}

The discovery of Fornax was presented in the same paper as the
discovery of Sculptor (Shapley 1938), and they make an interesting
pair for comparison.  Qualitatively they look very similar, although
Fornax is larger and more metal rich in the mean, and has globular
clusters, whereas Sculptor has none. Also, looking in detail at the
star formation histories of Sculptor and Fornax it is clear they have
followed very different evolutionary paths. Sculptor is dominated by
older stellar populations, whereas Fornax appears to have been forming
stars quite actively until 2~Gyr ago, and to be dominated by a
4$-$7~Gyr old population, and has evidence for only a small number
of globular cluster age stars in its field population.

Fornax contains a large number of RR~Lyr and Mira variables, and
carbon stars, as well as one anomalous Cepheid, one planetary nebula
and 5 globular clusters (Da Costa 1998). Young~(1999) looked for neutral
hydrogen, and found none.  The best estimate for the mean abundance of
the bulk of the stars in Fornax comes from the intrinsic width of the
RGB, and is [Fe/H]$= -$1.4 with a spread of about 0.15~dex (Buonanno
{\it et al.} 1985). It is quite hard to disentangle age from metallicity
effects on the RGB because Fornax has such a long lasting and complex
star formation history.

The extended sequence of main sequence turnoffs in the Fornax CMD
indicates a long history of star formation ({\it e.g.}, Beauchamp {\it
et al.} 1995; Stetson, Hesser \& Smecker-Hane 1998; Buonanno {\it et
al.} 1999), which has only recently ceased.  The luminosity of the
brightest blue stars show that Fornax cannot contain any stars younger
than 100~Myr (Stetson {\it et al.} 1998).

The presence of numerous ($\sim$ 120) carbon stars with a wide range
of bolometric luminosities indicates a significant mass dispersion
among the progenitors, and hence a significant age spread in the range
2-8 Gyr ago ({\it e.g.}, Azzopardi {\it et al.} 1999; Aaronson \&
Mould 1980, 1985).  This is supported by a well-populated
intermediate-age subgiant branch and a red clump, which require a
significant population with an age of 2-4~Gyr.  Most recently, an HST
study sampling the main-sequence turnoffs of the intermediate-age and
old populations in the centre of Fornax was carried out by Buonanno
{\it et al.} (1999), and they found evidence for a highly variable
star formation history starting at the epoch of globular cluster
formation (say 15 Gyr ago) and continuing until 0.5 Gyr ago. Their
detailed analysis did not take account of metallicity variations with
age, which must be present, which will probably make their ages
roughly 30\% too young, in the mean.

An old population is present as demonstrated by detection of a red
Horizontal Branch, slightly fainter than the Red Clump (Buonanno et
al. 1999), and of RR Lyrae variables (Stetson {\it et al.} 1998).  In
addition to the large number of RR~Lyrae variables, there is also a
weak, blue Horizontal Branch so Fornax clearly does contain a very
old, metal-poor component.

\subsubsection{Ca~II Triplet Results}

In Table~6 we summarise the results for the sample of stars we
observed in Fornax in a field in the centre of the galaxy, positioned
as shown in Figure~\ref{fnxcont}.  Where the observed stars lie in a
CMD is shown in Figure~\ref{fnxcmd}.  The Ca~II triplet results are
plotted on the equivalent width versus the V magnitude difference with
the Horizontal Branch, calibrated to the R97b results in
Figure~\ref{fnxres}.  Only stars which are considered to be members of
Fornax from their radial velocity, and lie on the RGB are fully
included in Figure~\ref{fnxres}; we have plotted the location (with
crosses) where the stars of low S/N lie, without their corresponding
error bars (see Table~6). This is done in the interests of not making
Figure~\ref{fnxres} unduly confusing.
Figure~\ref{fnxres} suggests that the bulk of stars in Fornax lie
between [Fe/H]$= -1.5$ and $-0.7$, with a mean value of $-1.0 \pm 0.3$.  
The total metallicity spread we
find for Fornax is 0.6~dex, and it is more skewed towards higher
metallicities than is consistent with that expected from the width of
the RGB.  There are apparently a few outliers at lower metallicity,
[Fe/H]$\sim -1.5$, and at higher metallicity [Fe/H]$\sim -0.5$.  The
abundance distribution is plotted as a histogram in
Figure~\ref{fnxfehist}. In contrast to Sculptor, Fornax has more of a
sharp cut-off in the metallicity distribution at low metallicities,
with a tail of values going out to higher metallicity values. The peak
of the distribution is at about [Fe/H]$= -$1.2.

As with Sculptor, the Red Giants selected across Fornax contain a
significant spread in metallicity, but not surprisingly the direct
comparison between our results and previous photometric determinations
of both the mean and the spread in metallicity do not agree.  The
metallicities we measure are higher and the spread is greater than the
photometric determinations. This is because Fornax contains a very
large spread in age and is dominated by intermediate age stars, and so
using globular cluster RGB fiducials is going to lead to incorrect
results, as the majority of Fornax stars are considerably younger than
globular cluster stars. This is why we find that the mean metallicity
of Fornax is higher than previously thought.  We do not have much
information on the spatial variation of metallicity in Fornax as both
our, relatively small, spectroscopic fields of view lie in the central
region of the galaxy, but over the region we do cover there is no
evidence for spatial variation in metallicity.

As for Sculptor, there is no correlation between the colour of a star
on the RGB and metallicity we measure from the Ca~II triplet lines in
Fornax.  The more metal rich and more metal poor populations over-lie
each other very closely.  But unlike Sculptor, the range of
metallicities and colours does match with the colour range of the
available isochrones at the metallicity of most of the stars. This
suggests that as we go to higher metallicities the stellar evolution
models do a better job on the RGB.

\subsubsection{An Evolutionary Scenario}

One possible global star formation history scenario for Fornax which
takes into account all the information that we now have on its stellar
population could be that, unlike Sculptor, Fornax appears to have
maintained some level of star formation more or less continuously over
most of its history until for the last few hundred million years. It
is also possible that the present gap in star-formation is temporary,
although it is then hard to understand where the gas for future
generations is now.  The main justification for this scenario comes
from the HST CMD analysis of (admittedly) a small fraction of the
stellar population of Fornax (Buonanno {\it et al.} 1999), and also
the apparently uniform distribution of Ca~II triplet metallicities
seen in our sample of RGB stars.

There is clear evidence in the evolved field stars of Fornax for very
old stars, from a blue Horizontal Branch and RR Lyrae stars, and also
an extensive intermediate age stellar population with a He burning Red
Clump, and Carbon stars. None of these indicators can be accurately
transformed into a star-formation rate at any of these times, except
perhaps that the unusually high fraction of Carbon stars in Fornax
suggests a large fraction of the stellar population of Fornax was
formed roughly between 1 and 9 Gyr ago, and the Red Clump suggests a
high star formation rate between 2 and 4 Gyr ago. These deductions can
be refined and confirmed by the HST CMD Main Sequence Turnoffs, which
suggest a dominant peak in star formation activity between 2.5 and 9
Gyr ago, with relatively very little star formation in the last
2.5~Gyr. It is complicated to quantify what may have happened more
than 9 Gyr ago. It seems from the Main Sequence Turnoff information
that more than 9~Gyr ago the star-formation rate was much lower than
during the peak period, 2.5$-$9~Gyr ago.  There is however the matter
of the 5 globular clusters to consider. Common wisdom says that
globular clusters accompany episodes of intense star formation in the
host galaxies ({\it e.g.}, Baade 1963).  This is however ambiguous
from the perspective of the dwarf galaxies of the Local Group ({\it
e.g.}, van den Bergh 2000), including Fornax.

Putting all the pieces of information we have about past star
formation in Fornax (but ignoring the presence of globular clusters),
one possible star formation history is plotted in the upper panel of
Figure~\ref{fnxsfh}.  Using this star formation history we estimate
what the accompanying metallicity evolution might have been, and this
is plotted as a dashed line in the lower panel of Figure~\ref{fnxsfh}.
We have assumed a simple chemical
evolution model (as first described by Searle \& Sargent 1972). 
We derived
the assumed yield (the rate at which stars are producing Fe) by 
taking as fixed the high metallicity end
point of the star formation history at the most recent time.
Also plotted on this lower diagram are where the RGB stars we observed
would lie if we determine ages, using isochrones (Bertelli et
al. 1994).  The trend for metallicity with age is clearly increasing
with time towards the present, but there is quite a lot of scatter,
especially for those stars which appear to be very young. This might
be due to problems in matching the low metallicity stellar evolution
tracks to our observations, which was also found to be a problem in
Sculptor.  As the metallicity of the stellar population gets lower the
difference in V$-$I colour of a 15~Gyr old star versus a 2~Gyr star on
the RGB in the theoretical models dramatically declines. At [Fe/H]$=
-0.7$ the difference is about 0.14~mag, but at [Fe/H]$=-1.9$ it is
about 0.04~mag. Therefore tiny reddening or photometric offset for low
metallicity stars will have a dramatic impact on the perceived age of
an RGB star.
There appears to be an offset in the basic metallicity
trend between the simple model
and the data points at ages $>$12~Gyr. This is the most uncertain
region to interpret the star formation history from the CMD. But as it
stands it might suggest that there has been some kind of initial
enrichment of the gas in Fornax at very early times. This could perhaps
be a relic of the epoch of globular cluster formation which appears to
pre-date that of the majority of star-formation in the field population
of Fornax.

%\newpage

\subsection{NGC~6822}

\subsubsection{Background}

NGC~6822 is one of our closest neighbouring dwarf-irregular (dI) type
galaxies. It was nominally discovered by Barnard (1884), although it
was not understood as neighbouring galaxy until Hubble (1925)
discovered Cepheid variable stars and determined it as being much more
distant than even the furthest globular clusters, and thus truly
extra-galactic. It is quite similar to the SMC, but it is
smaller and less luminous, and unlike the SMC, it is not involved in
any (obvious) interaction with our Galaxy or any other
galaxy. Unfortunately NGC~6822 is at very low Galactic latitude, and
therefore suffers from large and varied reddening over the whole
galaxy, and there is also a considerable problem of foreground stellar
contamination when accurately determining its stellar content. Despite
these issues, its proximity makes it a very rewarding object to study
the properties of young (relatively) low metallicity stars in an isolated
star-forming dwarf irregular galaxy.

Unlike Sculptor or Fornax, NGC~6822 is presently actively forming
stars.  It contains numerous, H~II regions, some extremely luminous,
spread over the face of the galaxy ({\it e.g.}, Hodge, Lee \&
Kennicutt 1988; O'Dell, Hodge \& Kennicutt 1999), and various studies
have catalogued the massive star content ({\it e.g.}, Westerlund {\it
et al.} 1983; Armandroff \& Massey 1985).  NGC~6822 has a large
extended halo of HI gas (de Blok \& Walter 2000; Brandenburg \&
Skillman 1998), going out well beyond the optical galaxy.  There have
been detailed abundance studies of young A-type super-giants (Venn
{\it et al.} 2000; Muschielok et al. 1999), and all the young stars
looked at to date reveal an iron abundance of [Fe/H]$\sim -0.50$,
confirming that NGC~6822 has a slightly higher present-day abundance
than the SMC, consistent with H~II region metallicities ({\it e.g.},
Pagel, Edmunds \& Smith 1980).  NGC~6822 also contains a number of
star cluster candidates (Wyder, Hodge \& Zucker 2000), although only
one remains as a likely ``true'' ancient globular cluster candidate,
similar to those found around our Galaxy.  Even one ancient globular
cluster is quite unusual for dwarf irregular galaxies in the Local
Group.

Cook, Aaronson \& Norris (1986) detected a significant population of
intermediate age carbon stars with a wide luminosity variation
(suggesting a large age range), and a significant extended AGB
population was detected by Gallart {\it et al.} (1994). The clear
presence of large numbers of these evolved stars shows that NGC~6822
must have had quite a high star formation rate during some or all of
the period 1-9~Gyr ago.  There has been an extensive series of papers
modelling the entire star formation history of NGC~6822 from the
resolved stars in a Colour-Magnitude Diagram by Gallart et al
(1996a,b,c). Although this work is extremely detailed and represents a
noble attempt to dig out information from a noisy CMD, their analysis
is based on data taken with a 2.5m telescope, and so the magnitude
limits are such that the main sequence turnoffs cannot be detected for
stars older than about 400~Myr, and so their results on star-formation
before this time come solely from the evolved red stellar population
of the upper RGB. This is subject to a great deal of uncertainty,
especially without any independent metallicity information. There is
little or nothing known about the stellar population older than about
9~Gyr in NGC~6822. The only CMD which reaches to the Horizontal Branch
luminosity comes from the analysis of field star contamination in the
HST studies of NGC~6822 star clusters, and has not yet been carefully
analysed in terms of a star formation history.  The Main Sequence in
the HST CMDs does not show any obvious evidence for strong or sharp
variations in the star formation rate with time over the past
Gyr. There is also only evidence for a very weak Horizontal Branch in
these data.  For the purposes of using the AD91 correction for the
effects of luminosity to the Ca~II triplet results, we estimated the
V$_{HB}$ from Wyder {\it et al.} to be 24.6.  
A search for RR Lyrae stars did not show them to be present in any 
significant numbers (Saha, private communication), and 
this study was contemporaneous with other searches by the 
same investigators, who successfully discovered RR Lyrae in several 
other galaxies, including IC 1613 (Saha {\it et al.} 1992).
This supports the HST
results, which covered only a small field of view, that the Horizontal
Branch, if present, is weak in NGC~6822.  This galaxy is clearly
dominated by intermediate age and perhaps even young stellar
populations.

NGC~6822 is nearly 3 magnitudes more distant than Fornax, and 4
magnitudes more distant than Sculptor. Thus these observations of
individual RGB stars are really pushing the capabilities of FORS1 
on UT1.

\subsubsection{Ca~II Triplet Results}

In Table~7 we summarise the results for the sample of stars we
observed in NGC~6822 in the NTT field observed at the position North
of the centre of the galaxy, shown in Figure~\ref{n6822cont}.  The
nomenclature for the stars in Table~7 is that stars with the prefix s1
refer to stars from the eastern half of the NTT field and s2 the
western half of the field. Those stars with s1b, refer to a second
series of observations on the eastern field, but for some reason they
were of much poorer signal-to-noise than the s1 data.

The Ca~II triplet results are plotted on the equivalent width versus
the V magnitude difference with the Horizontal Branch (see Table~1),
calibrated to the R97b results in Figure~\ref{n6822res}.  Only stars
which are considered to be members of NGC~6822 from their radial
velocity, and lie on the RGB are included in Figure~\ref{n6822res}; we
have plotted the location (with crosses) of where the particularly low
S/N stars lie, but without their error bars (see Table~7). Note that
there is a larger than usual uncertainty in the Horizontal Branch
magnitude, and there is almost certainly variable reddening across
NGC~6822. We have assumed a constant reddening (E(B$-$V)=0.26) for
all stars observed, but if there is indeed a large scatter about this
value this will artifically create a scatter in metallicity, but
this will almost certainly remain much less than
the intrinsic equivalent width measurement errors.

Although the error bars are large, it appears that the Red Giants
across the NGC~6822 field we observed contain a significant spread in
metallicity, which is larger than that seen in either Sculptor or
Fornax. The highest metallicity appears to be around [Fe/H]$= -0.5$,
and there are a few objects which fall around [Fe/H]$= -$2.  
The mean metallicity of the stars we observe is [Fe/H]$= -1 \pm 0.5$.
We don't
see any evidence for spatial variations, but we are observing a rather
small area of NGC~6822.  Our upper (young) metallicity measurements
are consistent with existing accurate young star and H~II regions
abundances for this galaxy, which consistently find [Fe/H]$= -0.5$.
NGC~6822 thus appears to contain stars across the entire metallicity
range to which we are sensitive ([Fe/H]$> -2.5$), unlike Fornax, which
appears to only contains star more metal rich than [Fe/H]$> -1.7$, or
Sculptor which only contains stars more metal poor than [Fe/H]$<
-1.3$.  The distribution of metallicities as a histogram
(Figure~\ref{n6822fehist}) also looks different to either Sculptor or
Fornax. It has a central peak at [Fe/H]$ = -$0.9 with a roughly equal
distribution of stars to the metal rich and metal poor side of this
peak.

\subsubsection{An Evolutionary Scenario}

The most striking aspect of Figures~\ref{n6822res} and
\ref{n6822fehist} is the large range in metallicity, but, because of
the problems with variable reddening and because the star formation
history of NGC~6822 is not constrained in detail beyond about 400~Myr
ago, it is more difficult to accurately fit the Ca~II triplet
observations into a detailed picture of the star formation history of
NGC~6822.  Thus, our attempt is much more speculative than for
Sculptor or Fornax.  Our uncertain knowledge about the detailed star
formation history is compounded by a combination of poorer S/N spectra
and variable and uncertain reddening towards NGC~6822.

Putting together what information we have results in a plausible star
formation history in the upper panel of Figure~\ref{n6822sfh}.  This
is derived from the results of Gallart {\it et al.} (1996a), the
carbon star survey of Cook {\it et al.} (1986), and the interpretation
of the extended AGB of Gallart {\it et al.} (1994) as being indicative
of a fairly recent ($\sim$~3~Gyr ago) burst of star formation (see
Lynds {\it et al.} 1998).  
In the lower panel we plot, as a dashed
line what might be a corresponding metallicity evolution, assuming the
star formation history given in the upper panel.  
We have assumed a simple chemical
evolution model (as first described by Searle \& Sargent 1972). 
We derived
the assumed yield (the rate at which stars are producing Fe) by 
taking as fixed the high metallicity end
point of the star formation history at the most recent time.
Also plotted on this
lower diagram are the RGB stars we observed, if we use isochrones
(Bertelli et al. 1994) to determine their ages. There were a large
number of stars in NGC~6822 which were significantly too red to match
any isochrone at their metallicity, presumably due to increased
reddening and these stars were not included in Figure~\ref{n6822sfh}.
Also plotted in the lower panel of Figure~\ref{n6822sfh} is the Venn
{\it et al.}~(2000) direct measurement of the iron abundance of two
(young) super-giant stars in NGC~6822; this is the present day
metallicity of the galaxy.
In Figure~\ref{n6822sfh} the significant rise in the star formation
rate of NGC~6822 which occurred at some point in the last 2$-$3~Gyr is
consistent with a relatively recent surge in the metallicity of
NGC~6822, as exhibited by the metallicity spread of the RGB stars in
this age range.  There are also a number of old, metal poor RGB stars,
which appear to come from a period before this recent increase in
star-formation activity.  Since the low end of the Ca~II triplet
metallicity is around [Fe/H]$= -2.4$, and there is little evidence for
a large old population, with only a very weak Horizontal Branch
population, there is unlikely to have been very much star formation
during ancient times in this galaxy.
It is hard to make a firm conclusion as to how good a match the data
points are to a simple chemical evolution model. 
Our results are broadly consistent with a trend of increasing metallicity with
time since formation over the last few Gyr. But the single 
point at 14~Gyr, if we chose to believe it, 
does suggest that the metallicity evolution
of this system has not been straight forward. But detailed analysis of
possible scenarios is premature with these sparse data. We await
more sensitive observations in the future.

\section{Conclusions}

We have presented spectroscopic metallicity distribution functions for
RGB stars in three nearby galaxies, that are more precise than
photometric estimates, and it is the first time that old-star
metallicity estimates have been made for NGC~6822.
We have attempted to reconcile photometric determinations
of star-formation histories from CMD analysis and spectroscopic
abundance determinations from Ca~II triplet equivalent widths into a
single star-formation and chemical evolution scenario over the period
1$-$15~Gyr ago for three nearby galaxies. 

All three galaxies we have studied have
clearly had very different evolutionary paths, both in star-formation
history and thus not surprisingly in metallicity evolution as well.
The CMD analysis and the Ca~II triplet results give broadly consistent
results, but there are some discrepancies which mean that studies like
this are worthwhile. The Ca~II triplet results allow us additional
insight in modelling the evolution of a galaxy in terms of the two
most fundamental parameters - star formation rates and chemical
evolution - we are thus measuring these parameters independently
through time, and thus accurately determining the population box of
each galaxy ({\it e.g.}, Hodge 1989).

The mean metallicity of the RGB stars we observed in our sample of
galaxies increases with the mass of the galaxy, as would be predicted
by theory ({\it e.g.}, Ferrara \& Tolstoy 2000).  Thus, Sculptor is
the least massive of the three galaxies (6$\times 10^6 \Msun$) and it
also has the lowest mean metallicity, [Fe/H]$= -1.6$. Fornax is about
ten times more massive (7$\times 10^7 \Msun$), and it
has a mean metallicity about three times greater, [Fe/H]$=
-1.1$. NGC~6822 is roughly another order of magnitude more massive 
(2$\times 10^9 \Msun$), and the metallicity of the RGB stars we
observed have a very large spread, but the mean [Fe/H]$= -0.9$, so a
factor of about 1.5 greater than Fornax.  The distribution of the
metallicities of the stars in each of the galaxies is very different,
as can be seen by comparing Figures~\ref{sclfehist}, \ref{fnxfehist}
and \ref{n6822fehist}.  Although we are observing a small sample of
the RGB stars in each galaxy, they were randomly selected across the
width of the RGB, and so we can conclude that the different
distribution of metallicity means that each galaxy has had a different
enrichment history, consistent with their different star formation
histories.

Sculptor is a galaxy where the star formation history was apparently
truncated about 4$-$5~Gyr ago, Fornax is a galaxy which had fairly
continuous star formation rate over most of its history until a few
hundred million years ago, and NGC~6822 has apparently enjoyed at
least one relatively recent enhancement in its star formation rate.
Thus all of these galaxies have experienced varying star-formation
rates over extended time periods.

One might speculate that Sculptor was formed from nearly pristine gas,
which was rapidly enriched in an ancient episode of star formation
which also formed a large fraction of the stars in the galaxy.
Star-formation then continued at slower rate until it was sharply
cut-off about 4$-$5~Gyr ago, when the metallicity of the stars being
formed at this point was [Fe/H]$= -1.2$. This would explain the sharp high
metallicity cut-off in Figure\ref{sclfehist}. Fornax on the other hand
looks more like a galaxy which formed most of its stars out of
pre-enriched gas, perhaps starting around 10~Gyr ago, at [Fe/H]$=
-1.3$, this explains the sharp low metallicity cut-off in
Figure~\ref{fnxfehist}.  It is possible that the gas out of which the
stars in the Fornax galaxy were formed was pre-enriched by the
formation of its globular cluster population about 15~Gyr ago, and
most of the star formation in the galaxy didn't begin until several Gyr
after this time.  The metallicity distribution of NGC~6822 looks like
a combination of both Sculptor and Fornax, suggesting that it formed
stars from very early times, but at a slow rate, which increased up to
a maximum quite recently, and it is now decreasing again. Probably if
gas were removed from NGC~6822 about 4$-$5~Gyr ago it would look
similar in most respects to a metal-rich Sculptor rather than Fornax,
and we can thus speculate that the main difference between Sculptor
and NGC~6822 is that NGC~6822 was massive enough to keep its gas up to
the present day (see Ferrara \& Tolstoy 2000). NGC~6822 is clearly a
galaxy which should be more carefully studied in the future.

\acknowledgements{
{\bf Acknowledgements:}
We thank Evan Skillman, Andrea Ferrara \& Kim Venn for useful
discussions, and the Paranal Observatory Science Operations staff for
the excellent support we received during this run, especially Thomas
Szeifert, Andreas Kaufer \& Chris Lidman.
}

\newpage
\def\x{\enspace}
\def\xx{\enspace\enspace}
\def\xxx{\enspace\enspace\enspace}
\def\xxxx{\enspace\enspace\enspace\enspace}
\def\xxxxx{\enspace\enspace\enspace\enspace\enspace}
\def\jref#1 #2 #3 #4 {{\par\noindent \hangindent=3em \hangafter=1
      \advance \rightskip by 5em #1, {\it#2}, {\bf#3}, {#4} \par}}
\def\ref#1{{\par\noindent \hangindent=3em \hangafter=1
      \advance \rightskip by 5em #1 \par}}
\def\endtable{\endgroup}
\def\tableheight{\vrule width 0pt height 8.5pt depth 3.5pt}
{\catcode`|=\active \catcode`&=\active
    \gdef\tabledelim{\catcode`|=\active \let|=\vbar
                     \catcode`&=\active \let&=\nobar} }
\def\table{\begingroup
    \def\twidth{\hsize}
    \def\tablewidth##1{\def\twidth{##1}}
    \def\defaultheight{\vrule width 0pt height 8.5pt depth 3.5pt}
    \def\heightdepth##1{\dimen0=##1
        \ifdim\dimen0>5pt
            \divide\dimen0 by 2 \advance\dimen0 by 2.5pt
            \dimen1=\dimen0 \advance\dimen1 by -5pt
            \vrule width 0pt height \the\dimen0  depth \the\dimen1
        \else  \divide\dimen0 by 2
            \vrule width 0pt height \the\dimen0  depth \the\dimen0 \fi}
    \def\spacing##1{\def\defaultheight{\heightdepth{##1}}}
    \def\nextheight##1{\noalign{\gdef\tableheight{\heightdepth{##1}}}}
    \def\end{\cr\noalign{\gdef\tableheight{\defaultheight}}}
    \def\zerowidth##1{\omit\hidewidth ##1 \hidewidth}
    \def\hline{\noalign{\hrule}}
    \def\skip##1{\noalign{\vskip##1}}
    \def\bskip##1{\noalign{\hbox to \twidth{\vrule height##1 depth 0pt \hfil
        \vrule height##1 depth 0pt}}}
    \def\header##1{\noalign{\hbox to \twidth{\hfil ##1 \unskip\hfil}}}
    \def\bheader##1{\noalign{\hbox to \twidth{\vrule\hfil ##1
        \unskip\hfil\vrule}}}
    \def\spanloop{\span\omit \advance\mscount by -1}
    \def\extend##1##2{\omit
        \mscount=##1 \multiply\mscount by 2 \advance\mscount by -1
        \loop\ifnum\mscount>1 \spanloop\repeat \ \hfil ##2 \unskip\hfil}
    \def\vbar{&\vrule&}
    \def\nobar{&&}
    \def\hdash##1{ \noalign{ \relax \gdef\tableheight{\heightdepth{0pt}}
        \toks0={} \count0=1 \count1=0 \putout##1\end
        \toks0=\expandafter{\the\toks0 &\end} \xdef\piggy{\the\toks0} }
        \piggy}
    \let\e=\expandafter
    \def\putspace{\ifnum\count0>1 \advance\count0 by -1
        \toks0=\e\e\e{\the\e\toks0\e&\e\multispan\e{\the\count0}\hfill}
        \fi \count0=0 }
     \def\putrule{\ifnum\count1>0 \advance\count1 by 1
        \toks0=\e\e\e{\the\e\toks0\e&\e\multispan\e{\the\count1}\leaders\hrule\
hfill}
        \fi \count1=0 }
    \def\putout##1{\ifx##1\end \putspace \putrule \let\next=\relax
        \else \let\next=\putout
            \ifx##1- \advance\count1 by 2 \putspace
            \else    \advance\count0 by 2 \putrule \fi \fi \next}   }
\def\tablespec#1{
    \def\vdimens{\noexpand\tableheight}
    \def\tabby{\tabskip=0pt plus100pt minus100pt}
    \def\r{&################\tabby&\hfil################\unskip}
    \def\c{&################\tabby&\hfil################\unskip\hfil}
    \def\l{&################\tabby&################\unskip\hfil}
    \edef\templ{\noexpand\vdimens ########\unskip  #1
         \unskip&########\tabskip=0pt&########\cr}
    \tabledelim
    \edef\body##1{ \vbox{
        \tabskip=0pt \offinterlineskip
        \halign to \twidth {\templ ##1}}} }

\centerline{\bf Table 1: The Galaxy Sample}
\vskip-1cm

$$
\table
\tablespec{\l\l\l\c\c\c\c\r\r\l\l}
\body{
\skip{0.06cm}
\hline
\skip{0.025cm}
\hline
\skip{.2cm}
& Object & l & b & V(HB) & (m-M)$_0$ & E(B$-$V) & M$_V$ & v$_r$ & $\sigma$ & type & ref &\end
&        &   &   &       &           &          &       &       &          &      &     &\end
\skip{.1cm}
\hline
\skip{.05cm}
\hline
\skip{.45cm}
& Sculptor&287.5 &$-$83.2 &20.35 &19.54$\pm$0.08&0.02$\pm$0.02& $-$11.1   &110 &6 &dSph & 1,2,6,8&\end      
& Fornax  &237.1 &$-$65.7 &21.50 & 20.70$\pm$0.12 &0.03$\pm$0.01 & $-$13.2&53 &10 &dSph & 3,4,6,9 &\end
& NGC~6822&25.3  &$-$18.4 &24.6:& 23.45$\pm$0.15 & 0.26$\pm$0.04&$-$15.2 & $-$49& - & dI  & 5,6,7 &\end
\skip{.25cm}
\hline
\skip{0.025cm}
\hline
}
\endtable
$$
%\vskip2cm
\noindent{1. Kunkel \& Demers 1977;}
2. Majewski {\it et al.} 1999;
3. Beauchamp {\it et al.} 1995;
4. Buonanno {\it et al.} 1999;
5. Gallart {\it et al.} 1996a;
6. Mateo 1998;
7. Wyder, Hodge \& Zucker 2000; 
8. Queloz, Dubath \& Pasquini 1995; 
9. Mateo {\it et al.} 1991

\newpage

\renewcommand{\thefootnote}{\fnsymbol{footnote}}

\centerline{\bf Table 2: Calibration Globular Clusters}
\vskip-1cm
$$
\table
\tablespec{\l\l\l\c\c\r\r\c}
\body{
\skip{0.06cm}
\hline
\skip{0.025cm}
\hline
\skip{.2cm}
& Object & l & b &[Fe/H] & E(B$-$V) & V(HB) & (m-M)$_V$ & v$_r$ &\end
&        &   &   &       &          &       &           & (km/s)&\end
\skip{.1cm}
\hline
\skip{.05cm}
\hline
\skip{.45cm}
& 47~Tuc & 305.9 & $-$44.9 & $-$0.71 & 0.04 & 14.06 & 13.37 & $-$18.7 &\end
& Pal~12 & 30.5 & $-$47.7 & $-$1.00 & 0.02 & 17.13 & 16.47 & $+$27.8  &\end
& Rup~106& 300.9 & 11.7 & $-$1.45 & 0.20 & 17.80 & 17.25 & $-$44.0    &\end
& M~15   & 65.0 & $-$27.3 & $-$2.15 & 0.10 & 15.83 & 15.37 & $-$107.3 &\end
\skip{.25cm}
\hline
\skip{0.025cm}
\hline
}
\endtable
$$
\newpage
\renewcommand{\thefootnote}{\fnsymbol{footnote}}

\centerline{\bf Table 3: The Observations}
\vskip-1cm
$$
\table
\tablespec{\l\l\l\r\c\c\c\l}
\body{
\skip{0.06cm}
\hline
\skip{0.025cm}
\hline
\skip{.2cm}
& Date & Begin & Object & Exptime & Airmass & DIMM\footnotemark[1]& Effective\footnotemark[2]& Comments &\end
&      & UT    &        & ~~(secs)&         & (arcsec)           &  (arcsec)               & &\end
\skip{.1cm}
\hline
\skip{.05cm}
\hline
\skip{.45cm}
& 18Aug99 & 06:03 & Scl-centre & 30   & 1.11 & 0.5                & & pre-image &\end
&         & 08:26 & Scl-c1     & 2$\times$1000 & 1.04 & 0.7  & 0.58 &           &\end
&	  & 09:15 & Scl-c2     & 2$\times$1200 & 1.11 & 0.85 & 0.62 & 	        &\end
& 18Aug99 & 06:07 & Scl-out2   & 30   & 1.11 & 0.5  & & pre-image &\end
& 19Aug99 & 06:46 & Scl-o2     & 1200 & 1.05 & 0.9  & 0.7 &           &\end
&         & 07:11 &            & 1000 & 1.02 & 0.8  &	  &      &\end
&         & 07:35 &            & 1200 & 1.01 & 0.7  &	  &      &\end
& 18Aug99 & 06:11 & Scl-out1   & 30   & 1.10 & 0.6  & & pre-image &\end
& 19Aug99 & 08:57 & Scl-o1     & 2$\times$2000 & 1.12 & 0.7  & 0.6 &      &\end
\skip{.1cm}
%\hline
\skip{.1cm}
& 18Aug99 & 10:02 & Fnx-centre & 60   & 1.02 & 0.8  & & pre-image &\end
& 20Aug99 & 06:04 & Fnx-c1     & 2$\times$2000 & 1.2 & 0.6  & 0.64 &      &\end
&         & 08:49 & Fnx-c2     & 2$\times$2000 & 1.02 & 0.9  & 0.68 &      &\end
\skip{.1cm}
%\hline
\skip{.1cm}
& 18Aug99 & 05:56 & N6822-centre & 60   & 1.46 & 0.6  & & pre-image &\end
&         & 01:57 & N6822-susi1  & 3$\times$2000 & 1.03 & 0.5  & 0.56 &      &\end
& 19Aug99 & 02:43 & N6822-susi2  & 3$\times$2000 & 1.08 & 0.65 & 0.6 &      &\end
& 20Aug99 & 00:05 & N6822-susi1b & 3$\times$2000 & 1.08 & 0.6 &  0.8 &     &\end
\skip{.1cm}
%\hline
\skip{.1cm}
& 17Aug99 & 23:35 & Rup~106 & 2$\times$300 & 1.68 & 0.7  & 1.08 &      &\end
\skip{.1cm}
%\hline
\skip{.1cm}
& 19Aug99 & 04:44 & M~15    & 2$\times$120 & 1.26 & 0.44 & 0.5 &      &\end
\skip{.1cm}
%\hline
\skip{.1cm}
& 19Aug99 & 08:11 & 47~Tuc-l5406 & 150 & 1.49 & 0.9  & 1.0 &      &\end
&         & 08:32 & 47~Tuc-l3512 & 2$\times$200 & 1.50 & 0.7 & 0.84 &      &\end
\skip{.1cm}
& 20Aug99 & 03:43 & Pal~12 & 2$\times$300 & 1.02 & 0.5  & 0.54 &      &\end
\skip{.25cm}
\hline
\skip{0.025cm}
\hline
}
\endtable
$$
\footnotetext[1]{ This is just an indication of the external seeing 
measured automatically by the seeing monitor (DIMM) on the mountain. 
Usually the seeing on the instrument is better than this.}
\footnotetext[2]{This is the effective seeing measured from the spatial extent of the
spectra on the {\it combined} images.}
%&18Aug99& 00:02 & N6822 & 1107 &  ??    & 0.8  & uncertain astrom. &\end
%&19Aug99&02:12&N6822-susi2&520&1.02&0.5&lost guide star &\end

\newpage

\hskip5cm {\bf Table 4: Calibrator Results}
%\vskip-1.cm
$$
\table
\tablespec{\l\l\l\c\r\r\r\c\l}
\body{
\skip{0.06cm}
\hline
\skip{0.025cm}
\hline
\skip{.2cm}
& Star & V & B$-$V & $\Sigma W_{ob}$ & v$_r$(ob)~~~~ & off~~~ & v$_r$(c)~~~ & v$_r$(pr) &Ref &\end
&      &   &       &                 &   (km/s)~~~   &  (km/s)&   (km/s)    &   (km/s)  &    &\end
\skip{.1cm}
\hline
\skip{.05cm}
\hline
\skip{.45cm}
& 47tuc-l3512           & 11.8 & 1.63& 6.04$\pm$0.05& $-$47.2$\pm$1.6& $+$20 &$-$27.2  &$-$21&1&\end
& 47tuc-l5406           & 12.8 & 1.30& 5.41$\pm$0.13& $-$48.8$\pm$1.8& $+$21 &$-$27.8  &$-$29&1&\end
\skip{.1cm}
\hline
\skip{.1cm}
&m15-38	                & 14.4 & 0.97& 2.09$\pm$0.10& $-$33.4$\pm$2.8& $-$82 &$-$115.4 &&2&\end
&m15-24	                & 13.6 & 1.14& 2.56$\pm$0.14& $-$39.6$\pm$2.5& $-$39 &$-$79.0  &&2&\end
&m15-58\footnotemark[2]	& 16.1 & 0.77& 2.57$\pm$0.13& $-$50.3$\pm$4.2& $-$58 &$-$108.3 &&2&\end
&m15-195\footnotemark[2]& 16.4 & 0.72& 2.44$\pm$0.45& $-$44.1$\pm$7.4& $-$62 &$-$103.1 &&2&\end
&m15-260                & 14.5 & 0.97& 2.22$\pm$0.15& $-$40.6$\pm$2.1& $-$69 &$-$109.6 &&2&\end
&m15-302\footnotemark[6]& 14.2 & 0.99& 2.78$\pm$0.33& $-$69.1$\pm$3.8& $-$31 &$-$100.1 &&2&\end
&m15-371& 13.6\footnotemark[1] & 0.91& 2.27$\pm$0.25& $-$39.5$\pm$3.9& $-$74 &$-$113.5 &&2&\end
&m15-459& 14.4\footnotemark[1] & 0.80& 2.16$\pm$0.11& $-$43.9$\pm$2.7& $-$37 &$-$79.7  &&2&\end
&m15-462& 13.1\footnotemark[1] & 0.99& 2.97$\pm$0.23& $-$28.5$\pm$2.6& $-$78 &$-$106.4 &&2&\end
\skip{.1cm}
\hline
\skip{.1cm}
&pal12-3111\footnotemark[6]& 17.16& 0.76& 4.89$\pm$0.42 &$-$24.3$\pm$4.2&$+$39 &14.7 &&3&\end
&pal12-S1  & 14.58& 1.58& 6.45$\pm$0.11 &$-$21.7$\pm$2.6 & $+$68 & 46.3 &30.6&3,4,5&\end
&pal12-3460& 16.69& 0.83& 4.77$\pm$0.18 &    4.7$\pm$1.7 & $+$10 & 14.7 &    &3    &\end
&pal12-1118& 14.79& 1.52& 6.26$\pm$0.07 &$-$13.5$\pm$1.9 & $+$21 &  7.5 &25.3&3,4,5&\end
&pal12-1128& 15.35& 1.26& 5.50$\pm$0.09 &   65.8$\pm$1.7 & $-$60 &  5.8 &27.1&3,4  &\end
&pal12-1329& 17.06& 0.79& 4.32$\pm$0.22 &   26.8$\pm$1.9 & $+$12 & 38.8 &    &3    &\end
&pal12-1305& 15.86& 1.10& 5.39$\pm$0.12 & $-$4.7$\pm$1.4 & $+$33 & 28.3 &30.9&3,4  &\end
&pal12-3328& 17.17& 0.74& 4.27$\pm$0.16 &   44.3$\pm$1.8 & $-$33 & 11.3 &    &3    &\end 
\skip{.1cm}
\hline
\skip{.1cm}
&rup106-1730\footnotemark[2]& 14.8& 0.65& 3.86$\pm$0.10&   13.2$\pm$2.2 & $-$30& $-$16.8&     &6&\end
&rup106-1614\footnotemark[3]& 14.7& 1.67& 4.74$\pm$0.05&$-$53.5$\pm$1.5 &      &        &$-$54&5,6,7&\end  
&rup106-1067\footnotemark[2]& 13.5& 0.91& 5.60$\pm$0.09&   66.5$\pm$2.1 & $-$62& $+$4.5 &     &6&\end  
&rup106-1092                & 13.9& 1.27& 5.44$\pm$0.05&$-$86.2$\pm$1.9 & $-$4 & $-$90.2&     &6&\end  
&rup106-580&14.1\footnotemark[1]  & 1.17& 5.28$\pm$0.07&$-$15.3$\pm$1.4 & $-$51& $-$66.3&     &6&\end  
&rup106-236&14.3\footnotemark[1]  & 1.58& 5.04$\pm$0.05&   54.9$\pm$2.1 & $-$57& $-$2.1 &     &6&\end
\skip{.1cm}
\hline
\skip{.05cm}
\hline
}
\footnotetext[2]{Not an RGB star, not included in Figure~11}
\footnotetext[6]{Large error bars, not included in Figure~11}
\footnotetext[1]{Photometry corrected for relative flux in the through-slit image}
\footnotetext[3]{Defined as radial velocity standard}
\endtable
$$
\noindent{Sources of photometry, star identification and previous radial velocity
measurements:} \\
1 Da Costa \& Armandroff 1986;
2 Buonanno et al. 1983;
3 Harris \& Canterna 1980; \\
4 AD91; 
5 Brown {\it et al.} 1997;
6 Buonanno {\it et al.} 1990;
7 Da Costa {\it et al.} 1992
\newpage

\hskip5cm {\bf Table 5: Sculptor Results}
$$
\table
\tablespec{\l\l\l\l\c\r\r\r}
\body{
\skip{0.06cm}
\hline
\skip{0.025cm}
\hline
\skip{.2cm}
& Star & I & V$-$I & W$_{8542} +$ W$_{8662}$ & [Fe/H] &v$_r$(meas)~~ & offset~& v$_r$(corr) &\end
%&      &   &       &                        &        &   (km/s)~~~    &  (km/s)&   (km/s)    &\end
\skip{.1cm}
\hline
\skip{.05cm}
\hline
\skip{.45cm}
& c1-56\footnotemark[1] & 18.2 & 0.86  & 3.35 $\pm$ 0.24 & & 130.6 $\pm$ 2.4 & $-$2  & 128.6 &\end  
& c1-70 & 17.7 & 1.12  & 4.26 $\pm$ 0.24 &$-$1.28$\pm$0.10 & 101.2 $\pm$ 2.2 & $+$26 & 127.2 &\end
& c1-85 & 18.6 & 0.95  & 3.73 $\pm$ 0.46 &$-$1.30$\pm$0.19 &  75.9 $\pm$ 2.9 & $+$13 &  88.9 &\end
& c1-68 & 19.0 & 0.92  & 3.07 $\pm$ 0.43 &$-$1.49$\pm$0.18 & 113.2 $\pm$ 3.2 & $+$6  & 119.2 &\end
& c1-101& 17.7 & 1.04  & 3.21 $\pm$ 0.20 &$-$1.76$\pm$0.08 &  95.4 $\pm$ 1.9 & $+$15 & 110.4 &\end
& c1-99\footnotemark[4] & 18.2 & 1.00  & 4.54 $\pm$ 0.27 & & 133.3 $\pm$ 1.9 & $+$73 & 206.3 &\end
& c1-43 & 18.3 & 0.96  & 3.29 $\pm$ 0.48 &$-$1.56$\pm$0.20 & 162.3 $\pm$ 3.9 & $-$54 & 108.3 &\end
& c1-55\footnotemark[1] & 18.7 & 0.80  & 3.29 $\pm$ 0.71 & & 146.7 $\pm$ 4.0 & $-$43 & 103.7 &\end
& c1-67\footnotemark[5] & 19.0 & 0.79  & 3.69 $\pm$ 0.89 & & 170.2 $\pm$ 4.3 & $-$16 & 154.2 &\end
& c1-76 & 18.5 & 0.95  & 3.60 $\pm$ 0.43 &$-$1.40$\pm$0.18 & 122.5 $\pm$ 3.8 & $-$10 & 112.5 &\end
& c1-81 & 16.0 & 1.29  & 2.98 $\pm$ 0.11 &$-$2.23$\pm$0.05 & 101.3 $\pm$ 1.5 & $-$31 &  70.3 &\end
& c1-46 & 17.6 & 1.11  & 4.60 $\pm$ 0.52 &$-$1.18$\pm$0.22 & 193.7 $\pm$ 4.9 & $-$108&  85.7 &\end
& c1-47 & 17.7 & 0.99  & 3.88 $\pm$ 0.22 &$-$1.47$\pm$0.09 & 107.0 $\pm$ 2.2 & $+$4  & 111.0 &\end
& c1-88 & 17.5 & 1.11  & 4.40 $\pm$ 0.23 &$-$1.28$\pm$0.10 &  75.7 $\pm$ 2.8 & $+$35 & 110.7 &\end
& c1-78 & 17.2 & 1.19  & 3.98 $\pm$ 0.19 &$-$1.51$\pm$0.08 & 105.3 $\pm$ 2.8 & $+$7  & 112.3 &\end
& c2-64 & 17.1 & 1.08  & 3.27 $\pm$ 0.16 &$-$1.87$\pm$0.07 & 133.9 $\pm$ 1.7 & $-$32 & 101.9 &\end
& c2-81 & 18.1 & 1.03  & 4.10 $\pm$ 0.30 &$-$1.27$\pm$0.13 & 152.5 $\pm$ 2.3 & $-$23 & 129.5 &\end
& c2-88 & 18.7 & 0.92  & 3.61 $\pm$ 0.48 &$-$1.34$\pm$0.20 & 104.1 $\pm$ 4.4 & $+$3  & 107.1 &\end
& c2-73\footnotemark[1] & 18.2 & 0.92  & 2.90 $\pm$ 0.44 & & 103.8 $\pm$ 4.5 & 0.    & 103.8 &\end
& c2-38 & 18.9 & 0.91  & 2.99 $\pm$ 0.61 &$-$1.56$\pm$0.26 & 125.2 $\pm$ 4.2 & $-$29 &  96.2 &\end
& c2-39\footnotemark[1] & 18.2 & 0.89  & 2.53 $\pm$ 0.33 & & 134.8 $\pm$ 4.9 & $-$11 & 123.8 &\end
& c2-72 & 18.7 & 0.91  & 3.27 $\pm$ 0.62 &$-$1.49$\pm$0.26 & 95.5  $\pm$ 4.2 & $-$20 &  75.5 &\end
& c2-52 & 18.8 & 0.91  & 3.44 $\pm$ 0.74 &$-$1.41$\pm$0.31 & 142.2 $\pm$ 5.2 & $-$44 &  98.2 &\end
& c2-27 & 19.0 & 0.95  & 4.71 $\pm$ 0.63 &$-$0.81$\pm$0.27 & 115.6 $\pm$ 7.0 & $-$34 &  81.6 &\end
& c2-60 & 19.0 & 0.93  & 3.97 $\pm$ 0.71 &$-$1.12$\pm$0.30 & 159.1 $\pm$ 4.4 & $-$49 & 110.1 &\end
& c2-86 & 18.8 & 0.95  & 4.36 $\pm$ 0.75 &$-$1.01$\pm$0.32 & 158.0 $\pm$ 4.9 & $-$41 & 117.0 &\end
& c2-82 & 18.2 & 1.04  & 3.08 $\pm$ 0.34 &$-$1.68$\pm$0.14 & 168.8 $\pm$ 3.5 & $-$58 & 110.8 &\end
& c2-53 & 19.0 & 0.92  & 3.67 $\pm$ 0.65 &$-$1.25$\pm$0.27 & 135.0 $\pm$ 5.5 & $-$54 &  81.0 &\end
& c2-85 & 17.3 & 1.26  & 3.99 $\pm$ 0.18 &$-$1.49$\pm$0.08 & 165.3 $\pm$ 1.5 & $-$63 & 102.3 &\end
& o1-1  & 17.8 & 0.96  & 1.92 $\pm$ 0.15 &$-$2.30$\pm$0.05 & 122.6 $\pm$ 2.8 & $-$3  & 119.6 &\end
& o1-4  & 17.4 & 1.04  & 4.56 $\pm$ 0.18 &$-$1.27$\pm$0.05 & 114.9 $\pm$ 2.0 & $-$7  & 107.9 &\end
& o1-11\footnotemark[1] & 19.0 & 0.79  & 3.57 $\pm$ 0.46 & &  77.5 $\pm$ 4.1 & $+$32 & 109.5 &\end
& o1-14\footnotemark[5] & 19.5 & 0.49  & 3.41 $\pm$ 0.75 & & 138.6 $\pm$ 4.6 & $-$9  & 129.6 &\end
& o1-6  & 18.2 & 0.99  & 3.76 $\pm$ 0.20 &$-$1.42$\pm$0.08 &  98.9 $\pm$ 2.1 & $+$2  & 100.9 &\end
& o1-15\footnotemark[4]&19.9& 0.19  & 6.32 $\pm$ 0.97 & & 173.9 $\pm$ 6.1 & $-$6 & 167.9 &\end
& o1-21 & 17.4 & 1.00  & 3.01 $\pm$ 0.22 &$-$1.94$\pm$0.09 & 121.4 $\pm$ 1.9 & $-$17 & 104.4 &\end
& o1-22 & 18.8 & 0.91  & 2.24 $\pm$ 0.29 &$-$1.90$\pm$0.12 & 153.4 $\pm$ 7.6 & $-$39 & 114.4 &\end
& o1-30\footnotemark[2] & 17.9 & 2.33  & 2.36 $\pm$ 0.15 & &  14.0 $\pm$ 4.2 & $-$32 & $-$18. &\end 
& o2-28\footnotemark[1] & 18.8 & 0.72  & 1.69 $\pm$ 0.81 & & 151.1 $\pm$ 6.7 & $-$38 & 113.1 &\end
& o2-25 & 18.0 & 0.95  & 3.80 $\pm$ 0.39 &$-$1.43$\pm$0.16 & 150.9 $\pm$ 3.1 & $-$43 & 107.3 &\end
& o2-33 & 19.0 & 0.89  & 2.80 $\pm$ 0.76 &$-$1.60$\pm$0.32 & 142.3 $\pm$ 3.3 & $-$8  & 134.3 &\end
& o2-38 & 18.7 & 0.93  & 3.86 $\pm$ 0.57 &$-$1.23$\pm$0.24 & 125.6 $\pm$ 2.8 & $-$10 & 115.6 &\end
& o2-46 & 18.4 & 0.96  & 2.90 $\pm$ 0.53 &$-$1.71$\pm$0.22 & 141.5 $\pm$ 3.1 & $-$14 & 127.5 &\end
& o2-44 & 17.9 & 0.99  & 3.83 $\pm$ 0.34 &$-$1.44$\pm$0.14 & 145.2 $\pm$ 4.2 & $-$43 & 102.2 &\end
\skip{.1cm}
\hline
\skip{.05cm}
\hline
}
\footnotetext[4]{Probably not a Sculptor member, from radial velocity, not included in Figure~12.}
\footnotetext[5]{Probably a Horizontal Branch star in Sculptor, not included in Figure~12.}
\footnotetext[1]{Too blue to be an RGB star, not included in Figure~12.}
\footnotetext[2]{This could be a Galactic M-star, certainly not a Sculptor member, not included in Figure~12.}
\endtable
$$
\\
\newpage

\hskip5cm {\bf Table 6: Fornax Results}
%\vskip-1.cm
$$
\table
\tablespec{\l\l\l\l\c\r\r\r}
\body{
\skip{0.06cm}
\hline
\skip{0.025cm}
\hline
\skip{.2cm}
& Star & I & V$-$I & W$_{8542} +$ W$_{8662}$ & [Fe/H] & v$_r$(meas)~~ & offset~& v$_r$(corr) &\end
&      &   &       &                         &        &   (km/s)~~~    &  (km/s)&   (km/s)    &\end
\skip{.1cm}
\hline
\skip{.05cm}
\hline
\skip{.45cm}
& c1-660\footnotemark[1] & 19.0 & 0.91  & 6.94 $\pm$ 0.09 & &  88.7 $\pm$ 1.7 & $-$41 & 47.7 &\end 
& c1-350\footnotemark[6] &19.9 &0.83 &4.63 $\pm$ 0.97 &$-$0.92$\pm$0.41 & 75.9 $\pm$ 4.6 & $-$13 & 62.9 &\end 
& c1-444 & 19.3 & 1.04  & 5.55 $\pm$ 0.64 &$-$0.63$\pm$0.31 &  94.1 $\pm$ 3.2 & $-$23 & 71.1 &\end 
& c1-371 & 18.9 & 1.09  & 4.41 $\pm$ 0.33 &$-$1.21$\pm$0.17 &  84.0 $\pm$ 2.3 & $-$30 & 54.0 &\end 
& c1-601 & 18.8 & 1.03  & 6.22 $\pm$ 0.26 &$-$0.50$\pm$0.15 &  81.3 $\pm$ 3.0 & $-$32 & 49.3 &\end 
& c1-628 & 19.3 & 1.00  & 5.52 $\pm$ 0.52 &$-$0.67$\pm$0.29 &  99.7 $\pm$ 2.7 & $-$49 & 50.7 &\end
& c1-564 & 19.3 & 1.05  & 4.46 $\pm$ 0.40 &$-$1.11$\pm$0.25 &  79.6 $\pm$ 4.5 & $-$29 & 50.6 &\end
& c1-433\footnotemark[1] &20.2 &0.78 &4.55 $\pm$ 1.10 & & 94.6 $\pm$ 10.6 & $-$34 & 60.6 &\end
& c1-365\footnotemark[1] &19.8 &0.79 &6.01 $\pm$ 0.83 & &  81.1 $\pm$ 4.0 & $-$28 & 53.1 &\end
& c1-122\footnotemark[6] &20.1 &0.96 &5.15 $\pm$ 0.89 & $-$0.62$\pm$0.37& 104.8 $\pm$ 7.7 & $-$49 & 55.8 &\end
& c1-125\footnotemark[6] &19.8 &0.96 &6.35 $\pm$ 1.40 & $-$0.19$\pm$0.59&  90.3 $\pm$ 4.5 & $-$48 & 42.3 &\end
& c1-214 & 18.4 & 1.14  & 4.64 $\pm$ 0.33 &$-$1.25$\pm$0.14 &  92.2 $\pm$ 2.8 & $-$40 & 52.2 &\end
& c1-360 & 18.6 & 1.14  & 5.85 $\pm$ 0.23 &$-$0.68$\pm$0.14 &  65.3 $\pm$ 2.6 & 0.    & 65.3 &\end
& c1-200 & 18.8 & 1.05  & 4.58 $\pm$ 0.31 &$-$1.18$\pm$0.19 & 107.1 $\pm$ 3.0 & $-$25 & 82.1 &\end
& c1-344 & 19.1 & 1.01  & 4.52 $\pm$ 0.68 &$-$1.14$\pm$0.27 &  89.3 $\pm$ 5.4 & $-$36 & 53.3 &\end
& c2-822 & 19.6 & 1.03  & 5.43 $\pm$ 0.58 &$-$0.61$\pm$0.35 &  82.9 $\pm$ 3.5 & $-$21 &  61.9 &\end
& c2-777 & 19.0 & 1.06  & 4.27 $\pm$ 0.44 &$-$1.25$\pm$0.21 &  43.4 $\pm$ 3.4 & 0.    &  43.4 &\end
& c2-838 & 16.6 & 1.63  & 5.39 $\pm$ 0.09 &$-$1.27$\pm$0.02 &   1.6 $\pm$ 2.6 & $+$40 &  41.6 &\end
& c2-828 & 17.9 & 1.20  & 3.98 $\pm$ 0.18 &$-$1.64$\pm$0.07 &  72.0 $\pm$ 1.9 & $-$6  &  66.0 &\end
& c2-702 & 18.4 & 1.03  & 5.68 $\pm$ 0.26 &$-$0.83$\pm$0.12 &  64.5 $\pm$ 2.4 & 0.    &  64.5 &\end
& c2-613 & 18.9 & 1.08  & 6.02 $\pm$ 0.48 &$-$0.55$\pm$0.20 &  19.5 $\pm$ 2.0 & $+$13 &  32.5 &\end
& c2-769 & 18.2 & 1.20  & 5.55 $\pm$ 0.20 &$-$0.89$\pm$0.10 &  42.5 $\pm$ 2.5 & $+$8  &  50.5 &\end
& c2-511\footnotemark[1] & 18.5 & 0.86  & 5.18 $\pm$ 0.25 & &  47.6 $\pm$ 2.0 & $+$23 &  70.6 &\end
& c2-623 & 18.9 & 1.14  & 4.85 $\pm$ 0.36 &$-$1.01$\pm$0.18 &  66.0 $\pm$ 2.9 & $-$6  &  60.0 &\end
& c2-388 & 19.2 & 1.10  & 4.95 $\pm$ 0.46 &$-$0.92$\pm$0.30 &  52.7 $\pm$ 2.9 & $-$12 &  40.7 &\end
& c2-384\footnotemark[6] &19.9 &0.92 &5.26 $\pm$ 0.96 &$-$0.63$\pm$0.40 & 58.0 $\pm$ 7.7 & $+$2  &  60.0 &\end
& c2-294\footnotemark[1] &20.0 &0.87 &4.74 $\pm$ 1.15 & & 41.1 $\pm$ 8.8 & $+$15 &  56.1 &\end
& c2-552 & 18.4 & 1.03  & 4.80 $\pm$ 0.29 &$-$1.19$\pm$0.13 &  24.7 $\pm$ 1.4 & $-$1  &  23.7 &\end
& c2-249\footnotemark[1] &19.7 &0.86 &5.05 $\pm$ 0.93 & & 54.7 $\pm$ 5.5 & $+$14 &  68.7 &\end
& c2-413 & 18.4 & 1.18  & 5.32 $\pm$ 0.20 &$-$0.95$\pm$0.14 &  23.3 $\pm$ 2.0 & $+$18 &  41.3 &\end
& c2-647 & 18.8 & 1.09  & 4.80 $\pm$ 0.30 &$-$1.08$\pm$0.20 &  29.8 $\pm$ 1.8 & $+$4  &  33.8 &\end
& c2-621 & 19.0 & 1.06  & 4.33 $\pm$ 0.49 &$-$1.22$\pm$0.20 &  45.0 $\pm$ 5.3 & $+$14 &  59.0 &\end
& c2-41	 & 19.0 & 1.06  & 3.63 $\pm$ 0.59 &$-$1.52$\pm$0.26 &  82.6 $\pm$ 2.8 & 0.    &  82.6 &\end
\skip{.1cm}
\hline
\skip{.05cm}
\hline
}
\footnotetext[1]{Too blue to be an RGB star, not included in Figure~15.}
\footnotetext[6]{Large error bars on equivalent width measurements, plotted as
crosses in Figure~15.}
\endtable
$$
\newpage
\hskip5cm {\bf Table 7: NGC~6822 Results}
%\vskip-1.cm
$$
\table
\tablespec{\l\l\l\l\c\r\r\r}
\body{
\skip{0.06cm}
\hline
\skip{0.025cm}
\hline
\skip{.2cm}
& Star & I & V$-$I & W$_{8542} +$ W$_{8662}$ & [Fe/H] & v$_r$(meas)~~ & offset~& v$_r$(corr) &\end
&      &   &       &                         &        &   (km/s)~~~    &  (km/s)&   (km/s)    &\end
\skip{.1cm}
\hline
\skip{.05cm}
\hline
\skip{.45cm}
& s1-309 & 20.6 & 1.77 &2.17$\pm$0.38 &$-$2.34$\pm$0.16 &$-$40.9$\pm$10.4 & $-$10 &  $-$50.9 &\end
& s1-186\footnotemark[6]&20.6 &1.44 &5.14$\pm$1.23 &$-$1.18$\pm$0.52 &$-$47.7$\pm$9.6 &$-$6&$-$53.7 &\end
& s1-212\footnotemark[6]&20.0 &1.91 &7.19$\pm$0.92 &$-$0.36$\pm$0.39 &$-$100.4$\pm$3.4 &$+$26 &$-$71.4 &\end
& s1-210 &20.1 &1.52 &5.75$\pm$0.71 &$-$1.04$\pm$0.59 &$-$90.4$\pm$4.1 & $+$17 &  $-$73.4 &\end
& s1-200 &20.2 &1.64 &5.02$\pm$0.51 &$-$1.29$\pm$0.43 &$-$95.9$\pm$4.7 & $+$52 &  $-$43.9 &\end
& s1-59  &20.2 &1.26 &5.03$\pm$0.56 &$-$1.40$\pm$0.47 &$-$41.2$\pm$6.3 & $+$1  &  $-$40.2 &\end
& s1-188 &20.5 &1.44&6.69$\pm$0.78 &$-$0.58$\pm$0.65 &$-$45.8$\pm$4.4 & $-$7  &  $-$52.8 &\end
& s1-153 &20.2 &1.73&5.96$\pm$0.71 &$-$0.87$\pm$0.60 &$-$75.9$\pm$3.7 & $-$5  &  $-$80.9 &\end
& s1-43  &20.7 &1.60 &5.97$\pm$0.54 &$-$0.77$\pm$0.45 &$-$53.3$\pm$6.8 & $+$11 &  $-$42.3 &\end
& s1-205\footnotemark[4]&20.1 &1.80 &3.26$\pm$1.01 & &$+$21.7$\pm$5.9 &$+$11 &$+$32.7&\end
& s1-204\footnotemark[6]&20.6 &1.45 &4.68$\pm$1.35 &$-$1.38$\pm$0.57 &$-$53.8$\pm$6.5 &$+$25 &$-$28.8 &\end
& s1-378\footnotemark[4]&20.2 &1.77 &5.15$\pm$0.91 & &$-$104.0$\pm$5.4&$-$23 &$-$127.0&\end
& s1-111 &20.1 &1.53 &4.81$\pm$0.28 &$-$1.43$\pm$0.12 &$+$29.8$\pm$2.0 & $-$57 &  $-$27.2 &\end
& s1b-280&20.3 &1.66 &3.66$\pm$0.75 &$-$1.82$\pm$0.32 &$+$0.7$\pm$12.6 & $-$31 &  $-$30.3 &\end
& s2-208 &20.4 &1.61 &7.18$\pm$0.96 &$-$0.34$\pm$0.40 &$-$82.8$\pm$6.2 &$-$4  & $-$86.8 &\end 
& s2-246 &19.9 &1.80 &6.52$\pm$0.48 &$-$0.70$\pm$0.20 &$-$72.2$\pm$4.4 & $+$4 &  $-$68.2 &\end 
& s2-263 &20.2 &1.32 &6.09$\pm$0.54 &$-$0.92$\pm$0.23 &$-$30.3$\pm$4.6 &  0    &  $-$30.7 &\end 
& s2-352 &20.3 &1.62 &5.65$\pm$0.80 &$-$1.01$\pm$0.34 &$-$50.7$\pm$4.2 & $-$27 &  $-$77.7 &\end 
& s2-354 &20.1 &1.74 &6.62$\pm$0.85 &$-$0.63$\pm$0.36 &$-$5.7$\pm$5.0 & $-$55 & $-$60.7 &\end 
& s2-250 &20.1 &1.29 &6.33$\pm$0.68 &$-$0.86$\pm$0.29 &$-$60.1$\pm$6.2 & $-$25 &$-$85.1 &\end 
& s2-271 &20.2 &1.73 &7.56$\pm$0.75 &$-$0.20$\pm$0.32 &$-$47.9$\pm$6.3 & $-$25 &$-$72.9 &\end 
& s2-142 &20.0 &1.63 &3.71$\pm$0.46 &$-$1.91$\pm$0.19 &$-$48.6$\pm$5.4 & $-$10 &  $-$59.6 &\end 
& s2-248 &20.1 &1.70 &6.22$\pm$0.83 &$-$0.81$\pm$0.35 &$-$40.1$\pm$6.5 & $-$25 &  $-$65.1 &\end 
& s2-117\footnotemark[6]&19.9 &1.38 &5.60$\pm$1.16 &$-$1.21$\pm$0.49 &$-$65.7$\pm$3.6& $+$26 &$-$39.7&\end 
& s2-199\footnotemark[4]&20.6 &1.59 &5.09$\pm$0.87 & &$-$76.0$\pm$7.5 &$-$39 &$-$115.0&\end
& s2-198 &20.1 &1.59 &6.52$\pm$0.47 &$-$0.72$\pm$0.20 & $-$17.2$\pm$5.3 & $-$38 &$-$55.2 &\end 
\skip{.1cm}
\hline
\skip{.05cm}
\hline
}
\footnotetext[4]{Probably not a NGC~6822 member, from radial velocity, not included in Figure~18.}
\footnotetext[6]{Large error bars on equivalent width measurements, error bars not included in Figure~18.}
\endtable
$$

\newpage								     
\clearpage

%Figure 1
\begin{figure}								     
\centerline{\hbox{\psfig{figure=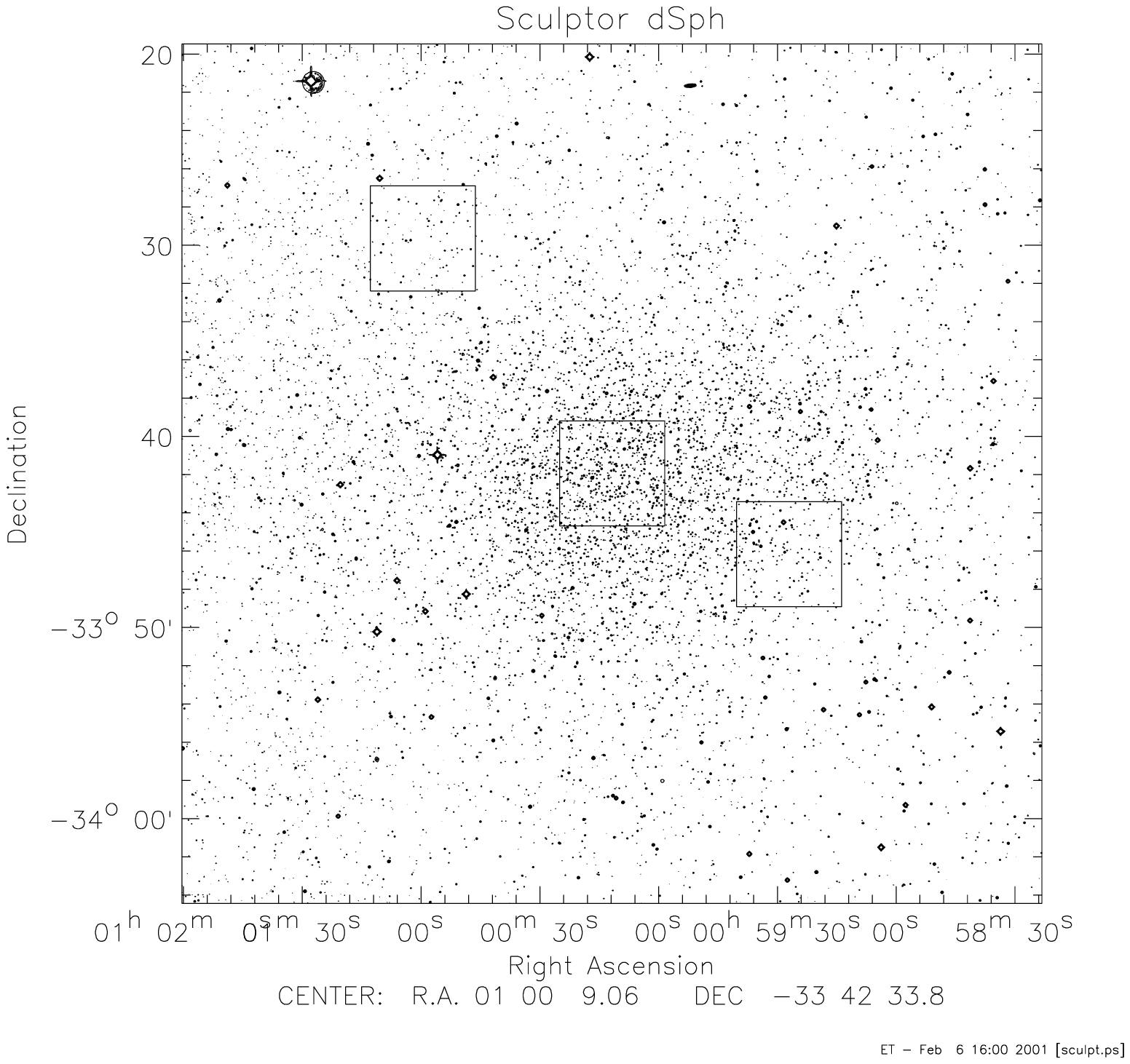,width=15cm}}}	     
\caption{
A contour plot 45$^{\prime}$ on a side of the Sculptor dSph galaxy, taken
from the Palomar Sky Survey. North is up and East is left.  The three
$5^{\prime}$ square fields for which we have NTT imaging, and thus from where
we have selected individual RGB stars, are shown.
}
\label{sclcont}								     
\end{figure}								     
									     
%Figure 2
\begin{figure}								     
\centerline{\hbox{\psfig{figure=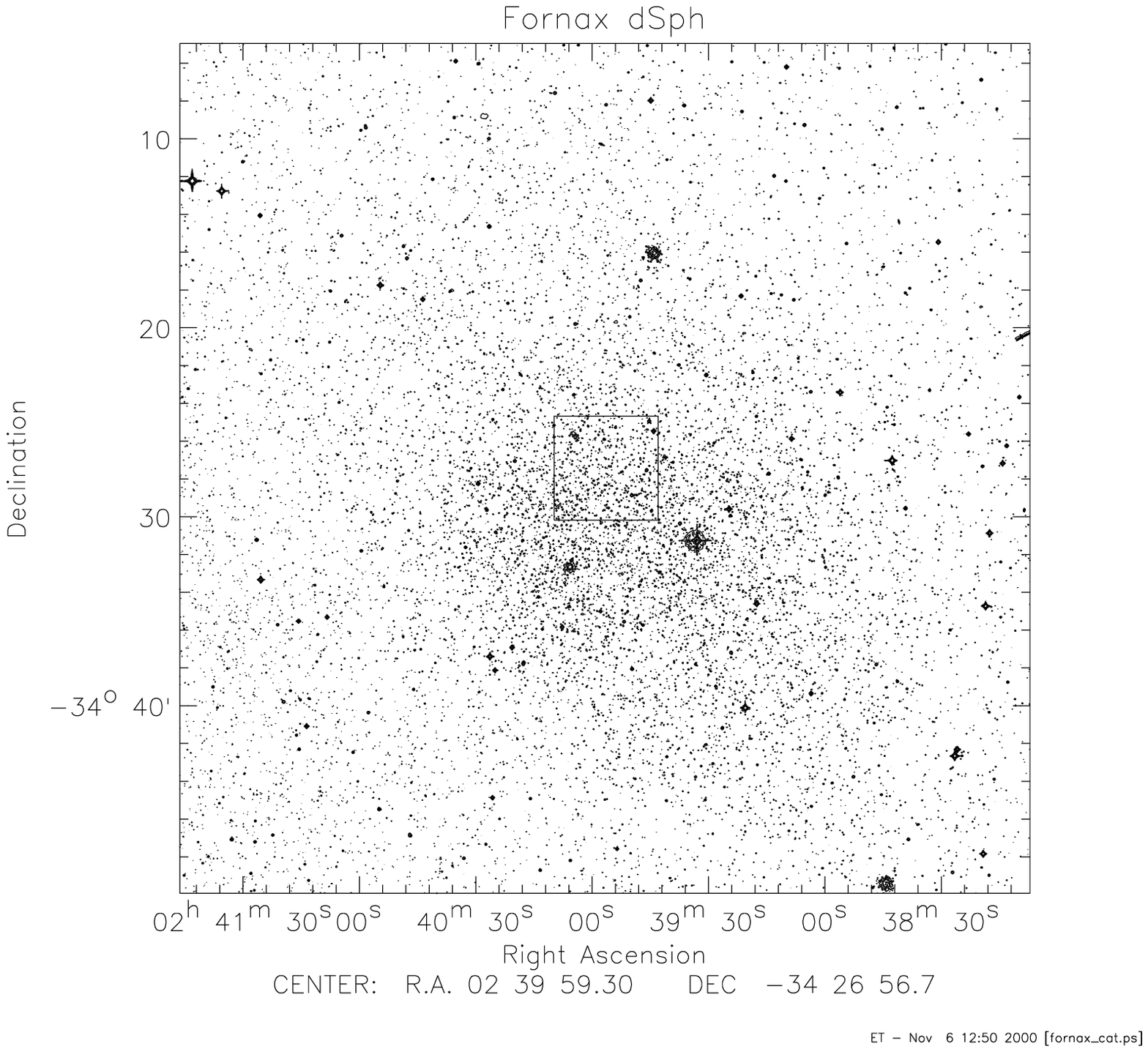,width=15cm}}}	     
\caption{
A contour plot 45$^{\prime}$ on a side of the Fornax dSph galaxy, same as Fig~1.
%from the Palomar Sky Survey. North is up and East is left.  The $5^{\prime}$
%square field for which we have NTT imaging, and thus from where we
%have selected individual RGB stars is shown.
}				     
\label{fnxcont}								     
\end{figure}								     
									     
%Figure 3
\begin{figure}								     
\centerline{\hbox{\psfig{figure=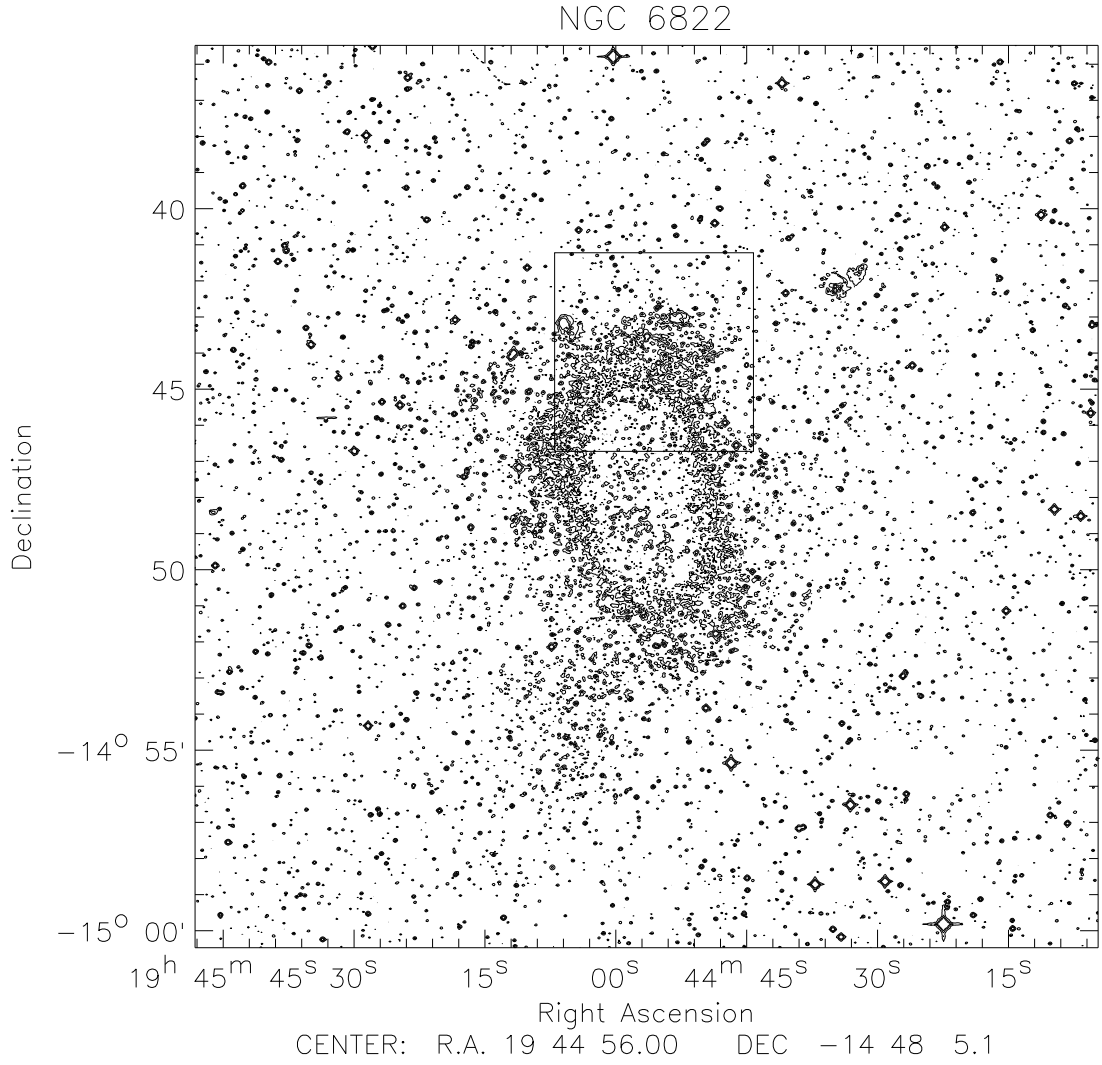,width=15cm}}}
\caption{
A contour plot 25$^{\prime}$ on a side of the NGC~6822 Dwarf Irregular
galaxy, same as Fig~1.
%taken from the Palomar Sky Survey. North is up and East is
%left.  The $5^{\prime}$ square field for which we have NTT imaging, and thus
%from where we have selected individual RGB stars is shown.
}
\label{n6822cont}
\end{figure}

%Figure 4
\begin{figure}
\centerline{\hbox{\psfig{figure=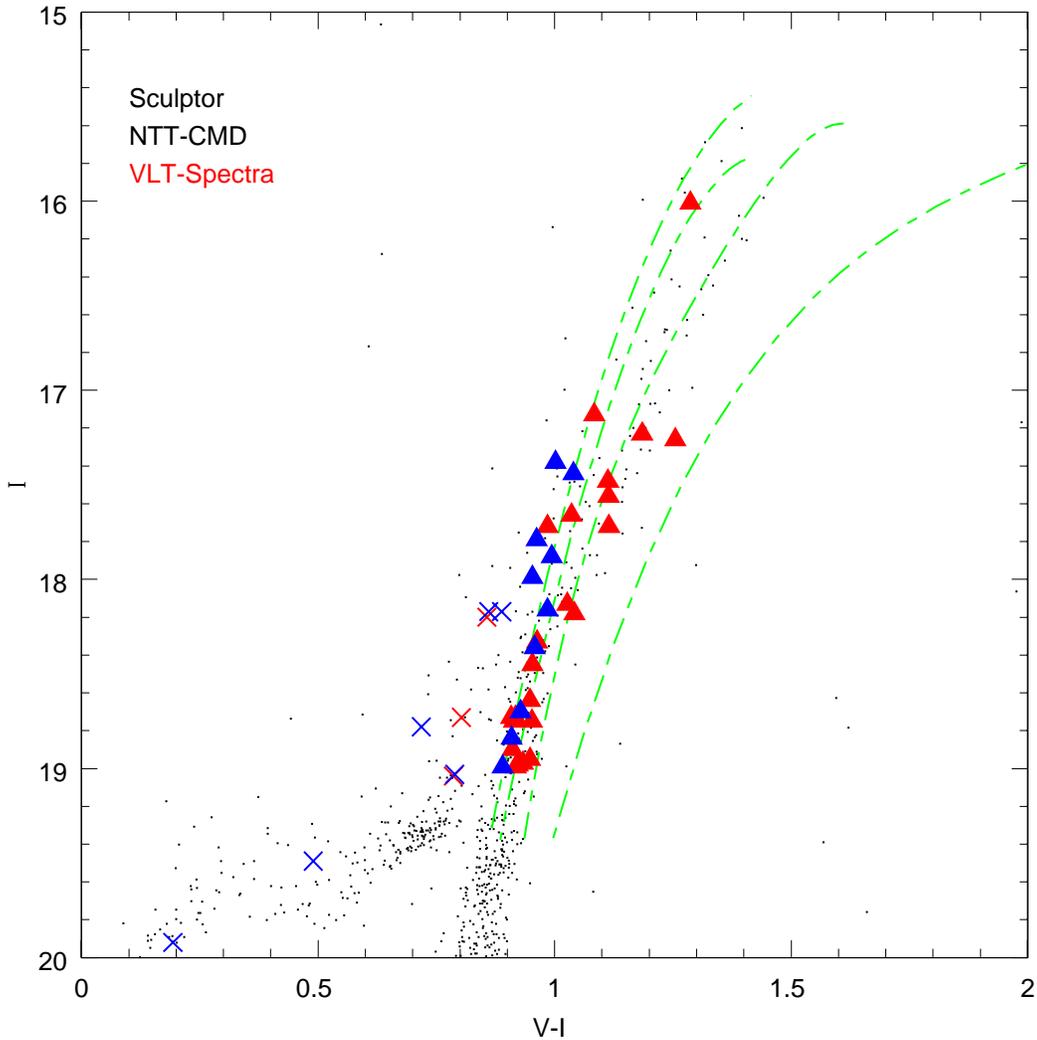,width=15cm}}}
\caption{
The combined NTT Colour-Magnitude Diagram of the stars in all three of
the Sculptor fields shown in Figure 1. Over-plotted in filled triangles
are the member stars for which we have FORS1 Ca~II triplet
spectroscopy.  The crosses are stars for which we have spectroscopy,
but either they are not radial velocity members, or the measurements
were not usable for one reason or another (see Table~5).  The dashed
lines are the RGB fiducials for the globular clusters: 47~Tucanae,
NGC~6752, NGC~6397 and M~15 from Da Costa \& Armandroff (1990), with
metallicities, [Fe/H]$= -0.7, -1.5, -1.9, -2.2$, respectively.
}
\label{sclcmd}
\end{figure}

%Figure 5
\begin{figure}
\centerline{\hbox{\psfig{figure=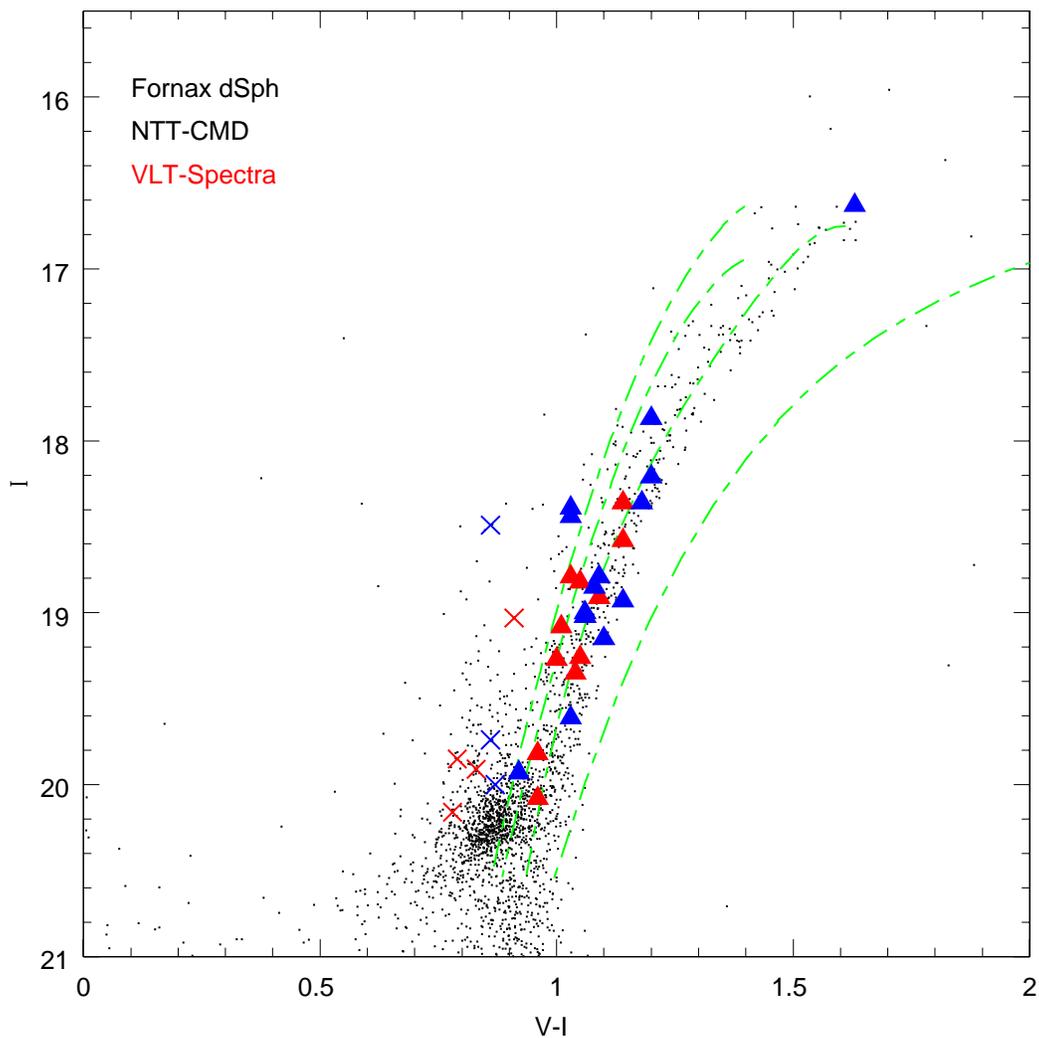,width=15cm}}}
\caption{
The NTT Colour-Magnitude Diagram of the stars in the Fornax field
shown in Figure 2. Over-plotted in filled triangles are the member
stars for which we have FORS1 Ca~II triplet spectroscopy.  The crosses
are stars for which we have spectroscopy, but either they are not
radial velocity members, or the measurements were not usable for one
reason or another (see Table~6).  The dashed lines are the RGB
fiducials for the globular clusters, see Fig~4.
%: 47~Tucanae, NGC~6752, NGC~6397
%and M~15 from Da Costa \& Armandroff (1990), with metallicities,
%[Fe/H]$= -0.7, -1.5, -1.9, -2.2$, respectively.
}
\label{fnxcmd}
\end{figure}

%Figure 6
\begin{figure}
\centerline{\hbox{\psfig{figure=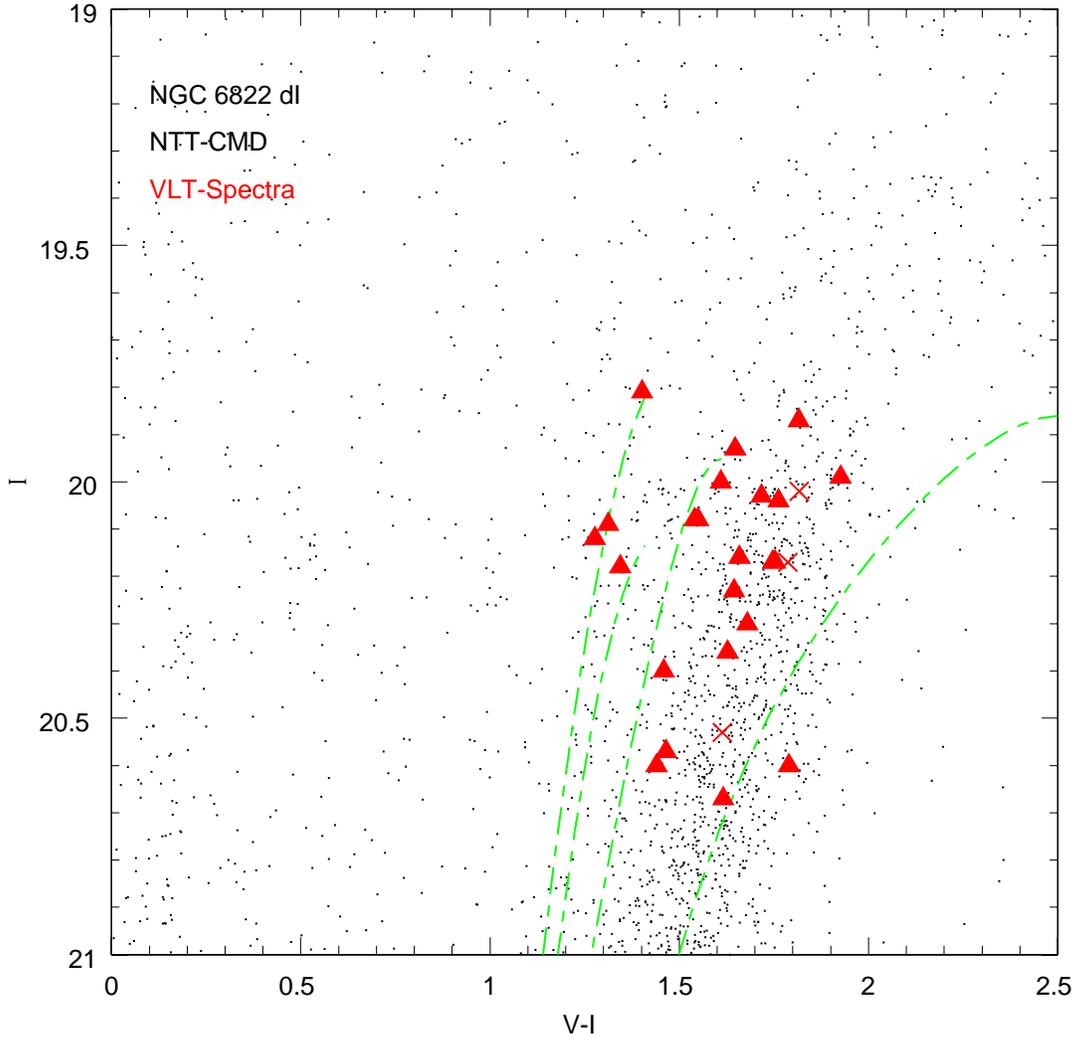,width=15cm}}}
\caption{
The NTT Colour-Magnitude Diagram of the stars in the NGC~6822 field
shown in Figure 3. Over-plotted in filled triangles are the stars for
which we have FORS1 Ca~II triplet spectroscopy, and which we believe
to be members based on the radial velocities. The crosses are stars
are not likely to be radial velocity members (see Table~7).  The
dashed lines are the RGB fiducials for the globular clusters, see Fig~4.
%47~Tucanae, NGC~6752, NGC~6397 and M~15 from Da Costa \& Armandroff
%(1990), with metallicities, [Fe/H]$= -0.7, -1.5, -1.9, -2.2$,
%respectively.
}
\label{n6822cmd}
\end{figure}

%Figure 7
\begin{figure}
\centerline{\hbox{\psfig{figure=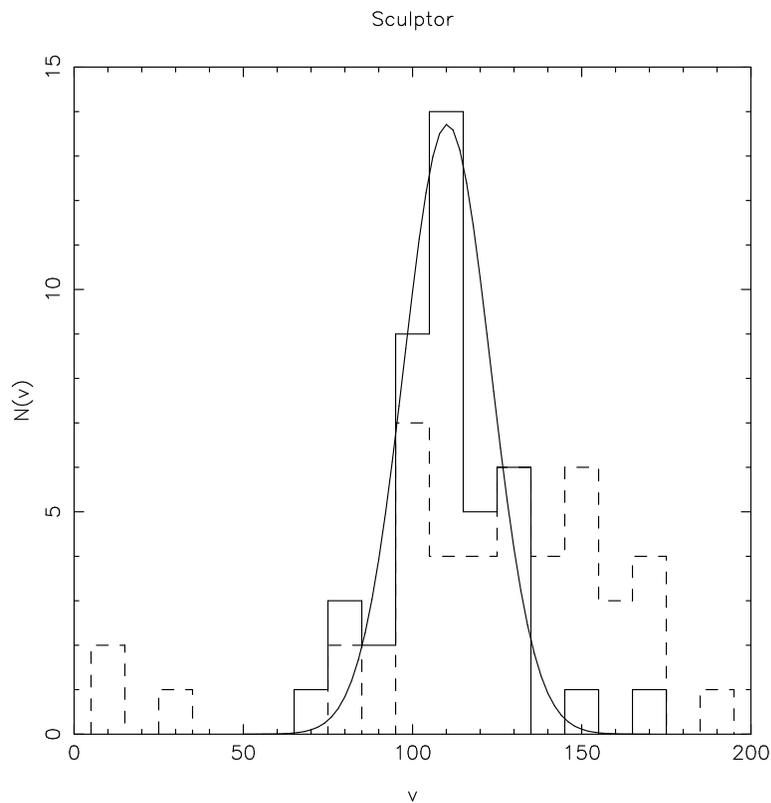,width=15cm,angle=-90}}}
\caption{A histogram of the radial velocity determinations for the
spectroscopically observed stars in Sculptor dSph (as shown in
Figure~4).  The dashed line represents the distribution of directly
measured radial velocities, and the solid line is the distribution
after corrections have been made for the position of each object in
the slit, with a Gaussian fit to these corrected points. The central
velocity we found here is 110.2 km/s, with a dispersion, $\sigma_v =
12.8$, which compares to the literature values of 110 km/s and
$\sigma_v = 6$ (Armandroff \& Da Costa 1986; Queloz {\it et al.} 1995),
giving a resulting accuracy of our velocity measurements at $\pm 10.7$
km/s.}
\label{sclhis}
\end{figure}

%Figure 8
\begin{figure}
\centerline{\hbox{\psfig{figure=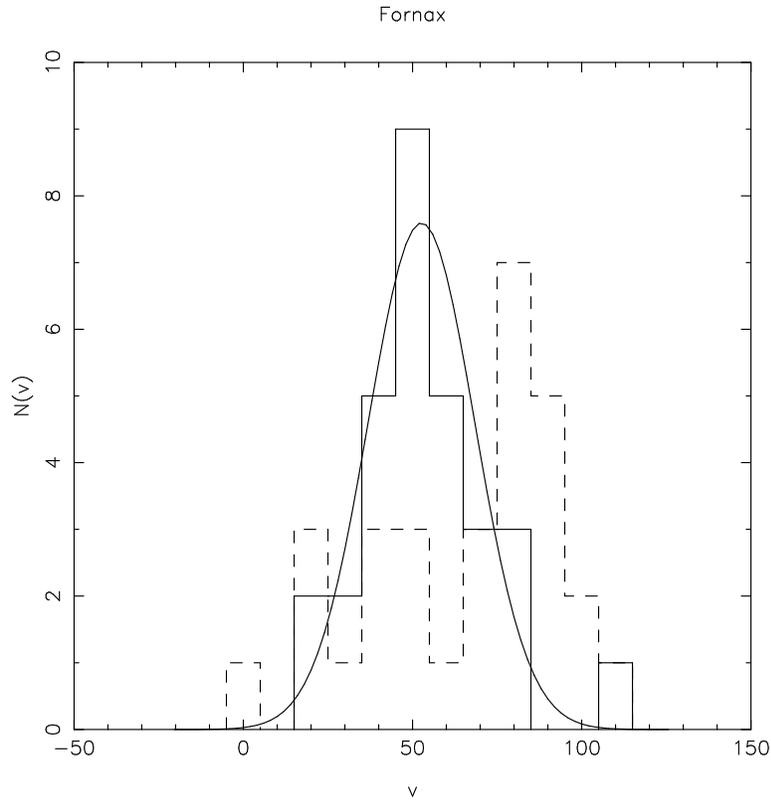,width=15cm,angle=-90}}}
\caption{A histogram of the radial velocity determinations for the
spectroscopically observed stars in Fornax dSph (as shown in
Figure~5).  The different lines are as described in Fig~7.
%The dashed line represents the distribution of directly
%measured radial velocities, and the solid line is the distribution
%after corrections have been made for the position of each object in
%the slit, with a Gaussian fit to these corrected points. 
The central
velocity we found here is 52.7 km/s, with a dispersion, $\sigma_v =
15.8$, which compares to the literature values of 53 km/s and
$\sigma_v = 11$ (Mateo {\it et al.} 1991), giving a resulting accuracy of
our velocity measurements at $\pm 11.3$ km/s.}
\label{fnxhis}
\end{figure}

%Figure 9
\begin{figure}
\centerline{\hbox{\psfig{figure=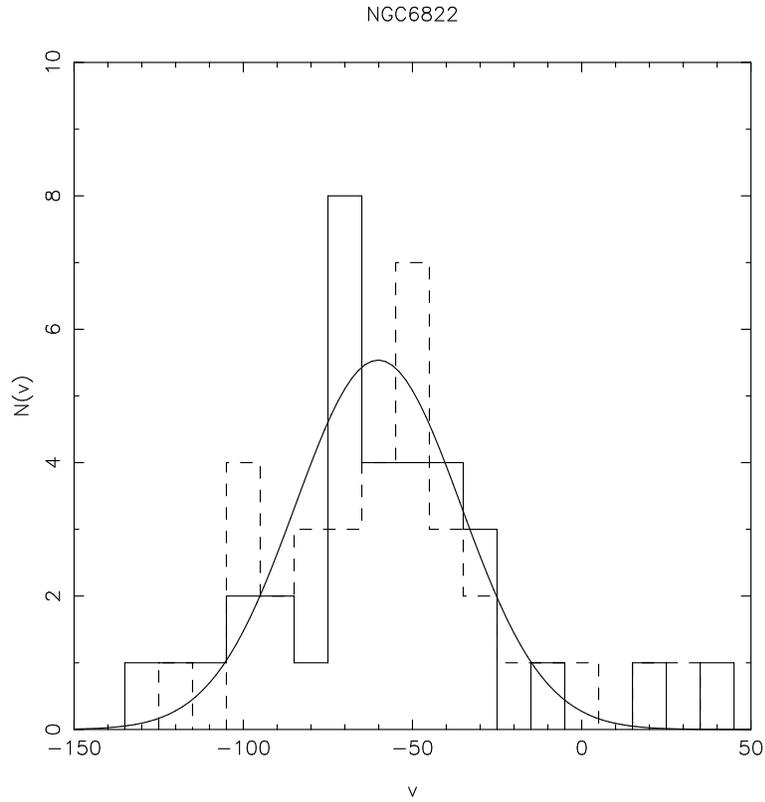,width=15cm,angle=-90}}}
\caption{A histogram of the radial velocity determinations for the
spectroscopically observed stars in NGC~6822 dI (as shown in
Figure~6).  The different lines are as described in Fig~7.
%The dashed line represents the distribution of directly
%measured radial velocities, and the solid line is the distribution
%after corrections have been made for the position of each object in
%the slit, with a Gaussian fit to these corrected points. 
The central
velocity we found here is $-$60.1 km/s, with a dispersion, $\sigma_v =
24.5$, which compares to the literature value of $-$57 km/s for the
central velocity (Richter, Tammann \& Huchtmeier 1987). The equivalent 
global velocity dispersion for N6822 from HI measurements is 34 km/s.
}
\label{n6822his}
\end{figure}

%Figure 10
\begin{figure}
\centerline{\hbox{\psfig{figure=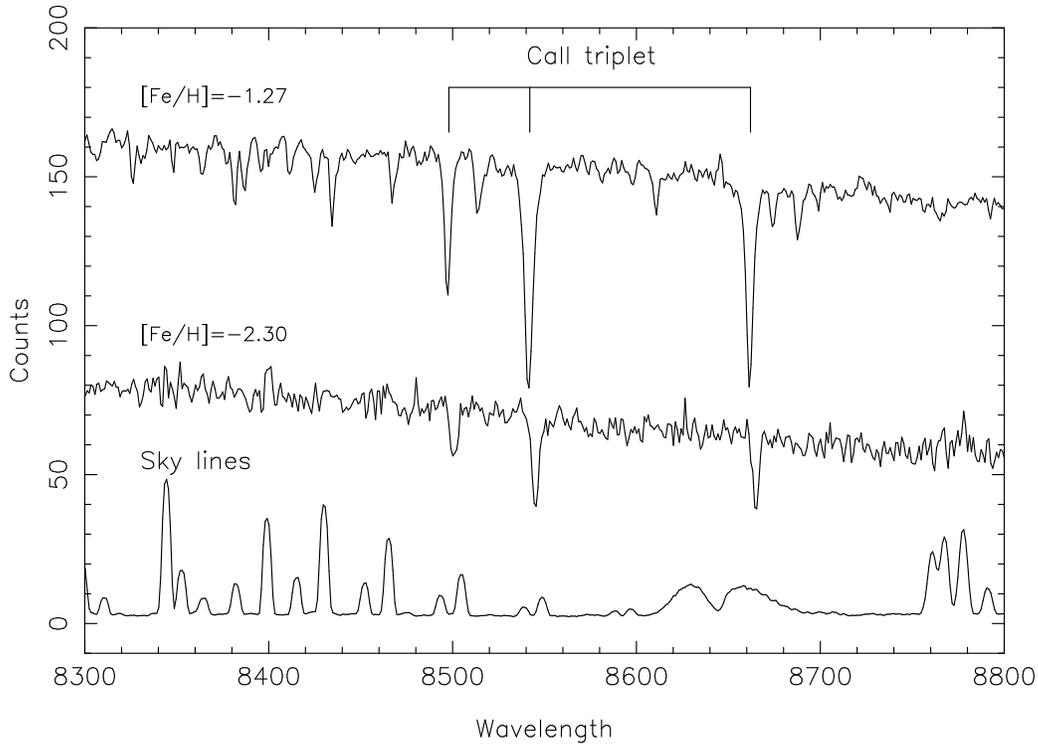,width=15cm,angle=-90}}}
\caption{
Here we show two example spectra at the opposite extremes of our
observed Ca~II triplet line widths.  For display purposes the spectra
have been normalized to their continuum level and then arbitrarily
shifted.  The upper spectrum is of star c2-838 in Fornax with a
calcium triplet metallicity of [Fe/H]$= -1.27$, and the lower spectrum
is of star o1-1 in Sculptor, with [Fe/H]$= -2.30$.  They both have
good S/N, with $\sim$30 in the upper spectrum and $\sim$20 in the
lower.  Also shown here is the sky spectrum. This shows that, although
this region of the spectrum is relatively free of bright sky lines,
the weaker Ca~II triplet line at 8498\AA \, is more likely to be
affected by sky lines than the other two.
}
\label{spec}
\end{figure}

%Figure 11
\begin{figure}
\centerline{\hbox{\psfig{figure=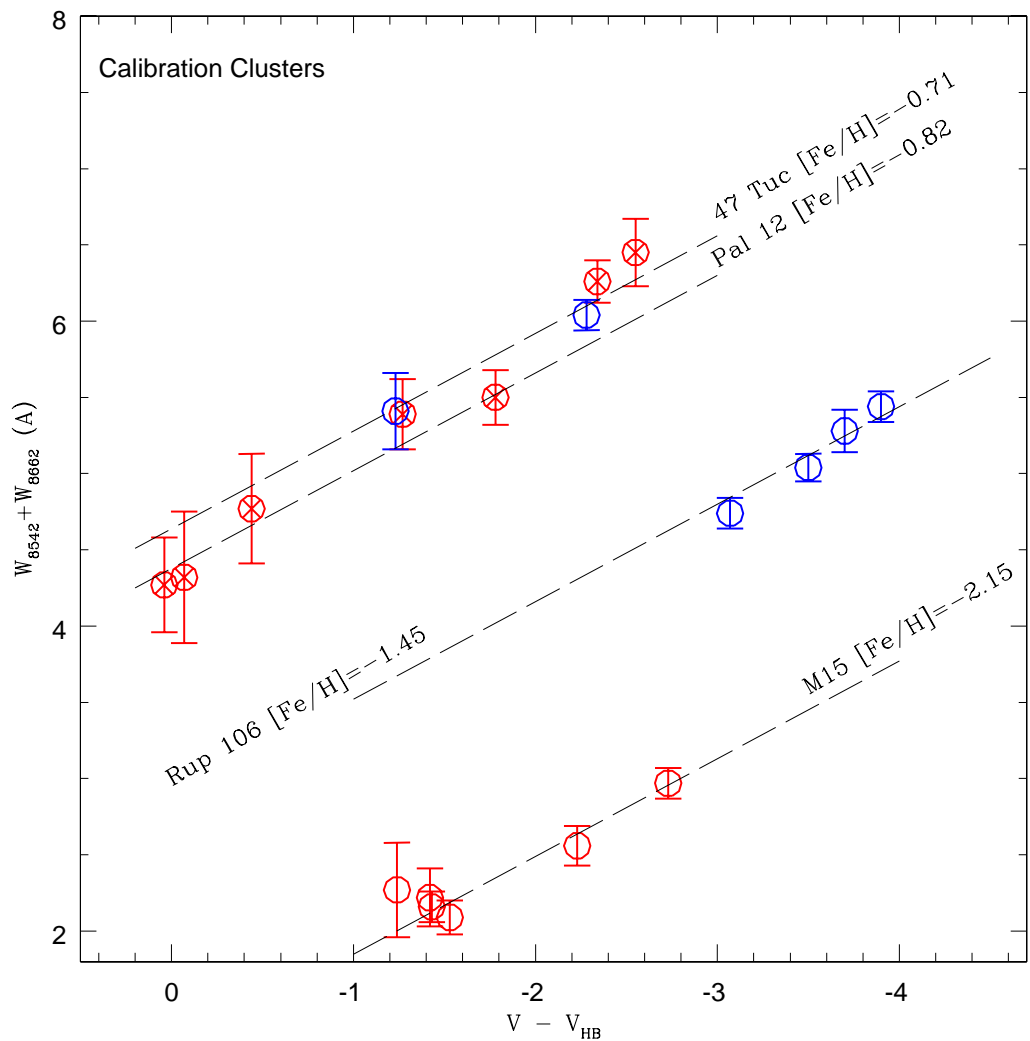,width=15cm}}}
\caption{
The summed equivalent width of the two stronger Ca~II triplet lines
(W$_{8542} +$ W$_{8662}$) is plotted against the V magnitude
difference with the Horizontal Branch (V $-$ V$_{HB}$) for the
observed stars in the calibration globular clusters Pal~12, 47~Tucanae,
Ruprecht~106 and M~15. The Pal~12 measurements are plotted as open
circles with crosses to distinguish them from the two 47~Tuc
measurements (open circles).  Also plotted for each cluster are the
best-fit lines, with slope 0.64 (see \S 3.2), and the metallicity
these correspond to is labeled.
}
\label{globs}
\end{figure}

%Figure 12
\begin{figure}
\centerline{\hbox{\psfig{figure=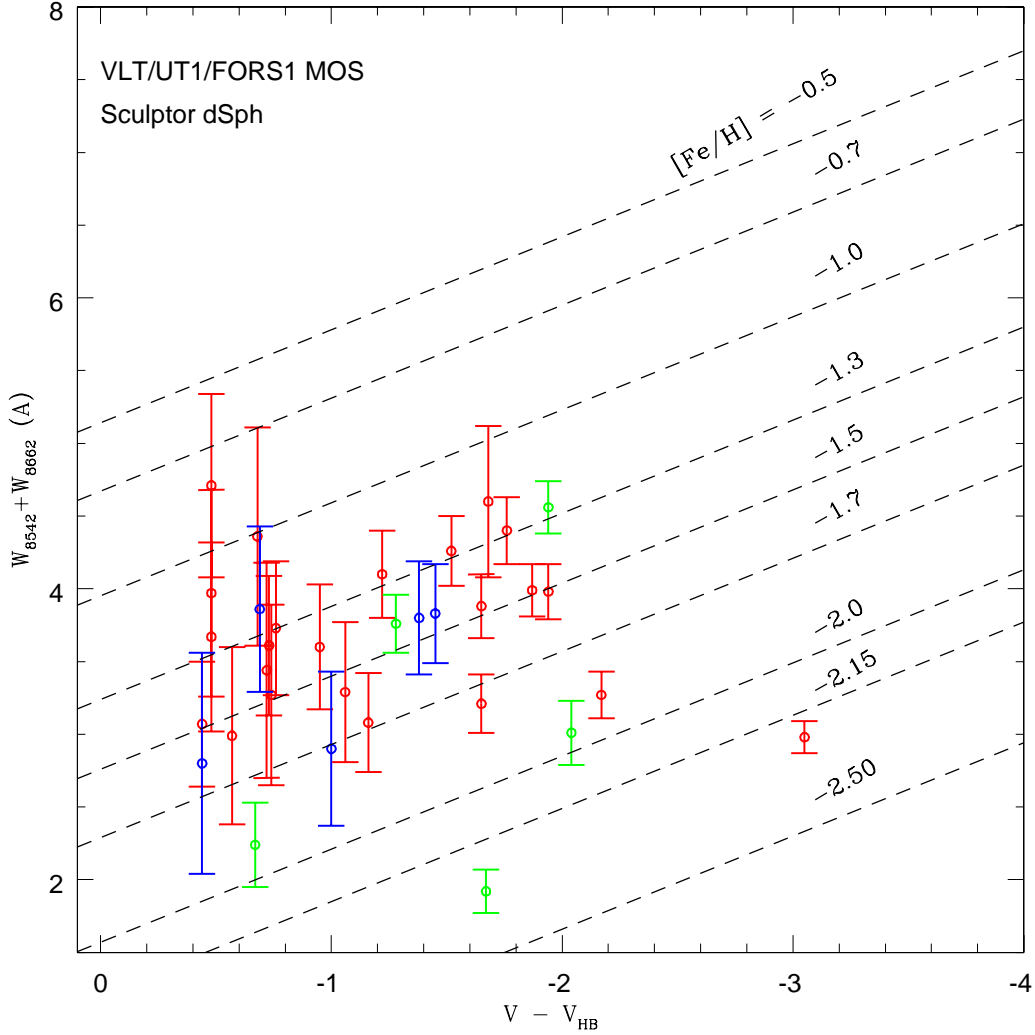,width=15cm}}}
\caption{
The summed equivalent width of the two stronger Ca~II triplet lines is
plotted against the V magnitude difference with the Horizontal Branch
for the observed stars in the Sculptor dSph.  Also plotted are lines
of constant metallicity, as calibrated and checked with R97, in Figure
10 (see \S 3.2 and \S 4.1).  The error bars come from the Gaussian
fitting measurement errors.
}
\label{sclres}
\end{figure}

%Figure 13
\begin{figure}
\centerline{\hbox{\psfig{figure=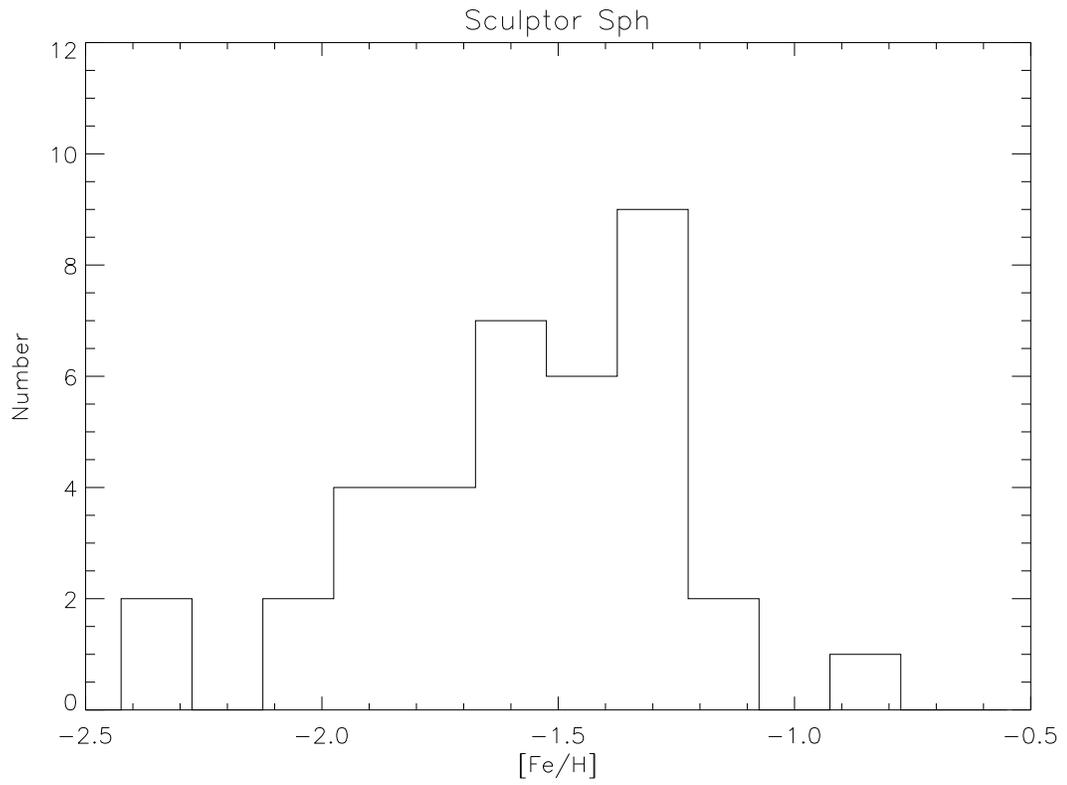,width=15cm}}}
\caption{
Here we plot the histogram distribution of Sculptor RGB Ca~II triplet
metallicities.
}
\label{sclfehist}
\end{figure}

%Figure 14
\begin{figure}
\centerline{\hbox{\psfig{figure=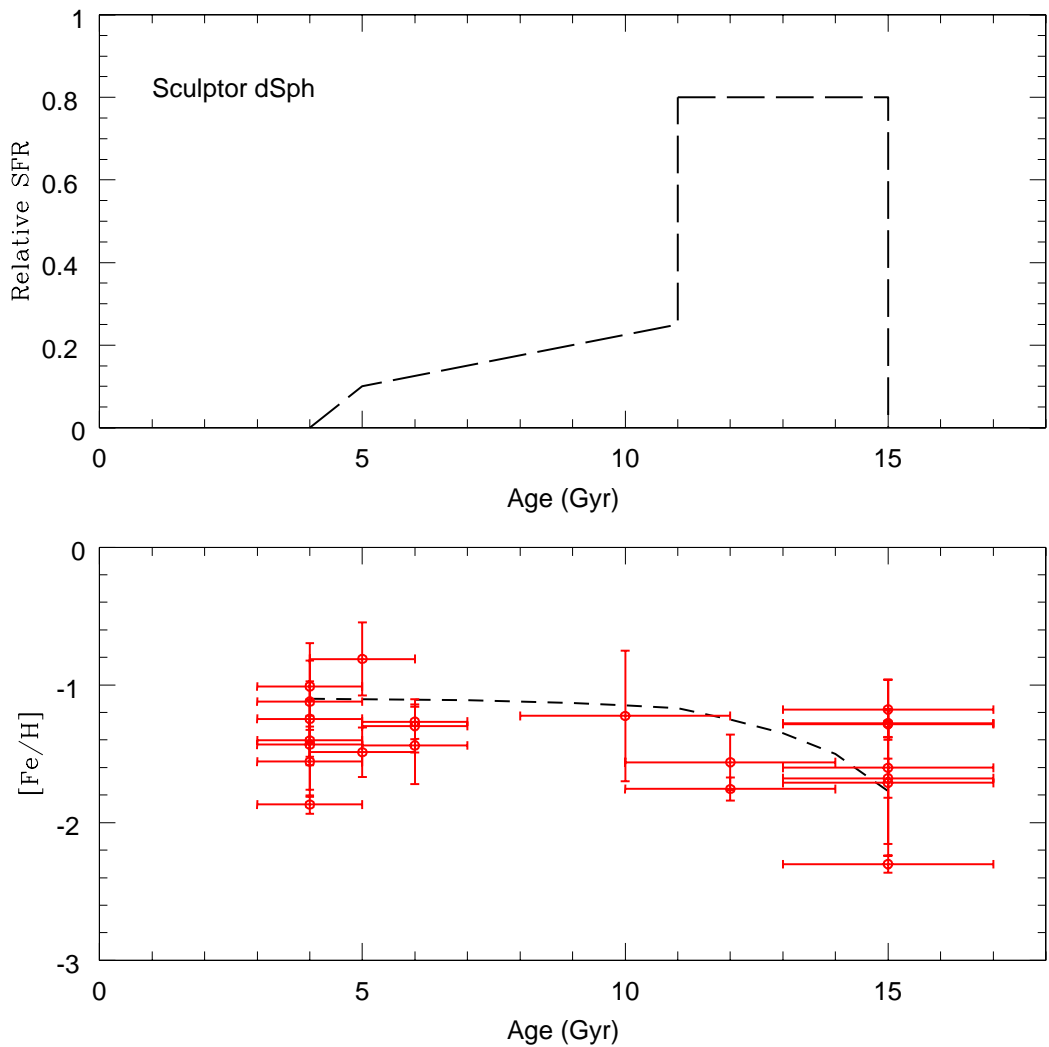,width=15cm}}}
\caption{
Here we display a possible star-formation and chemical evolution
scenario for Sculptor over its entire history ($\sim$15~Gyr).  In the
upper panel is a schematic plot, consistent with all that we know of
the stellar population of Sculptor, of how the rate of star formation
may have varied over time, back from the epoch of globular cluster
formation around 15~Gyr ago. In the lower panel we plot a
corresponding variation in metallicity over the same time frame.
Overploted on the lower panel are our Ca~II triplet measurements for
individual RGB stars, for which we determined ages using isochrones.
}
\label{sclsfh}
\end{figure}

%Figure 15
\begin{figure}
\centerline{\hbox{\psfig{figure=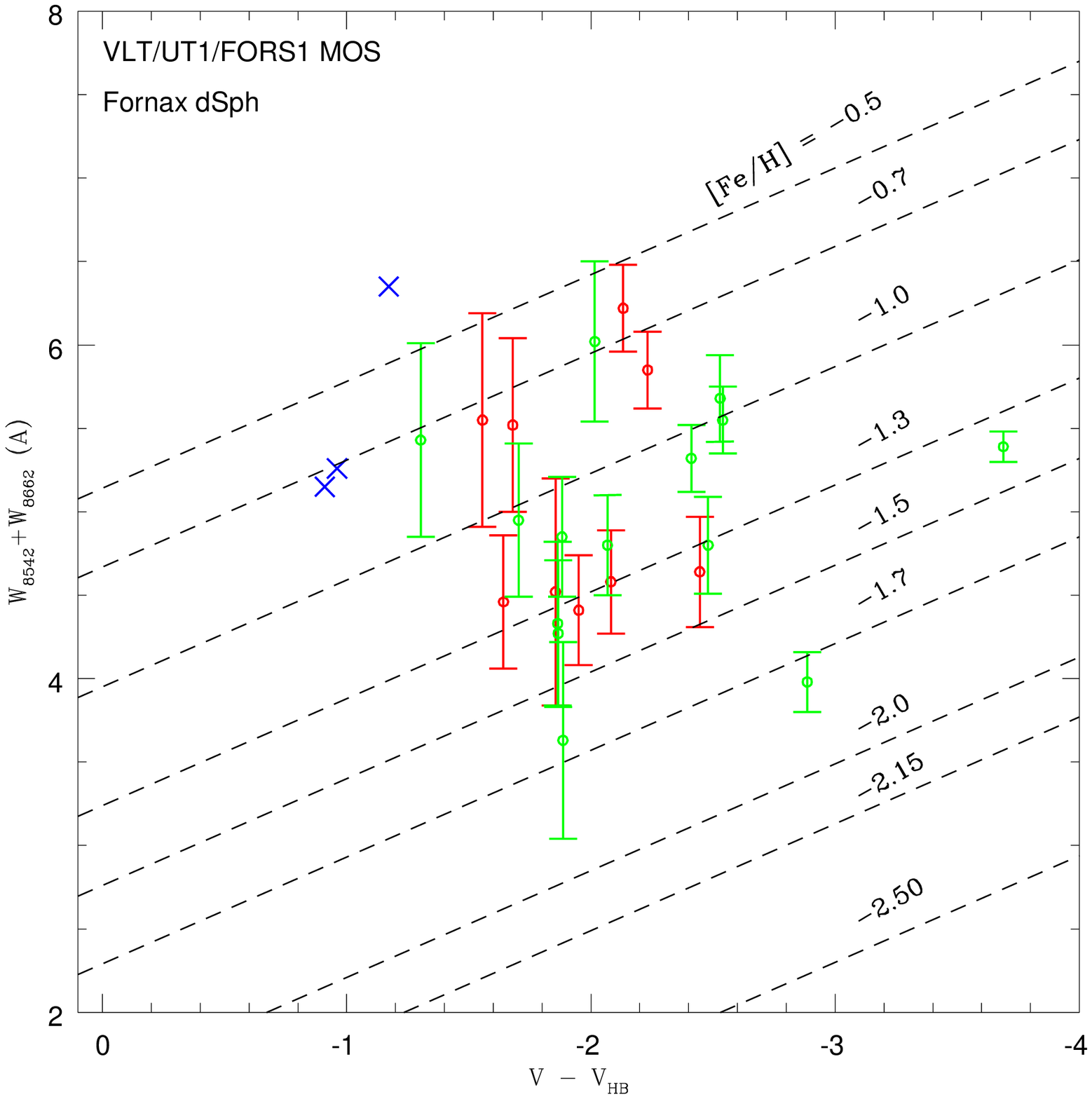,width=15cm}}}
\caption{
The summed equivalent width of the two stronger Ca~II triplet lines is
plotted, as described in Fig~12.
%against the V magnitude difference with the Horizontal Branch
%for the observed stars in the Fornax dSph.  Also plotted are lines of
%constant metallicity, as calibrated and checked with R97, in Figure 10
%(see \S 3.2 and \S 4.1).  The error bars come from the Gaussian
%fitting measurement errors.
}
\label{fnxres}
\end{figure}

%Figure 16
\begin{figure}
\centerline{\hbox{\psfig{figure=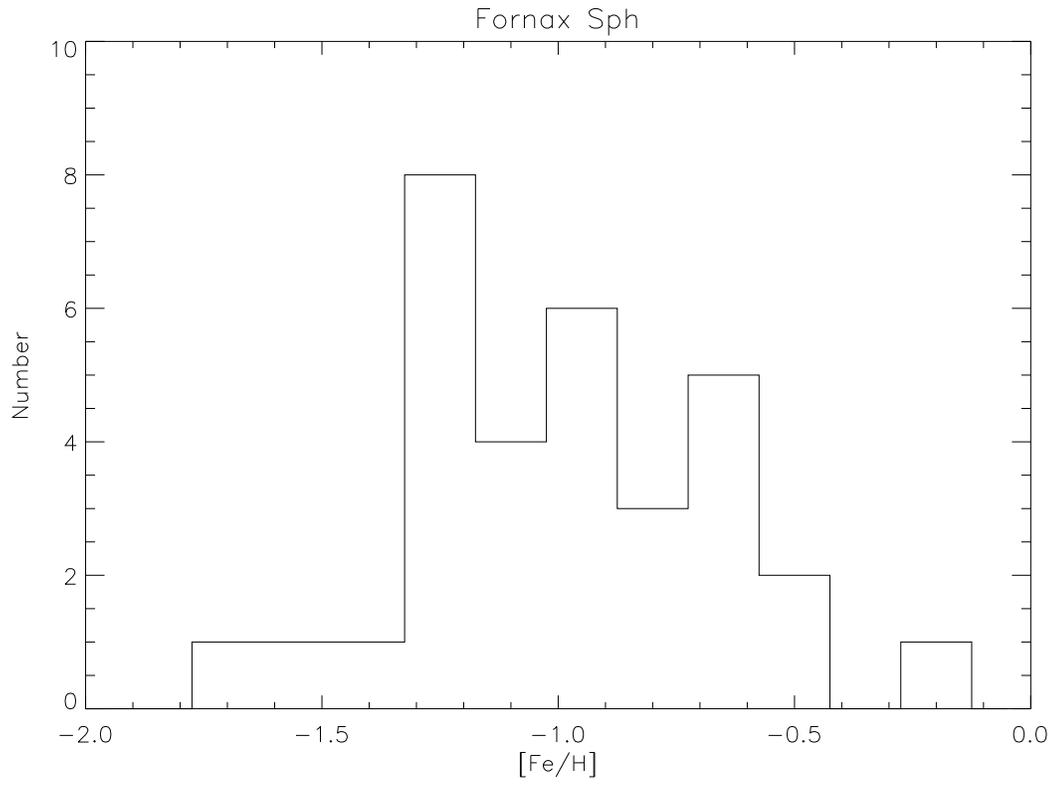,width=15cm}}}
\caption{
Here we plot the histogram distribution of Fornax RGB Ca~II triplet
metallicities.
}
\label{fnxfehist}
\end{figure}

%Figure 17
\begin{figure}
\centerline{\hbox{\psfig{figure=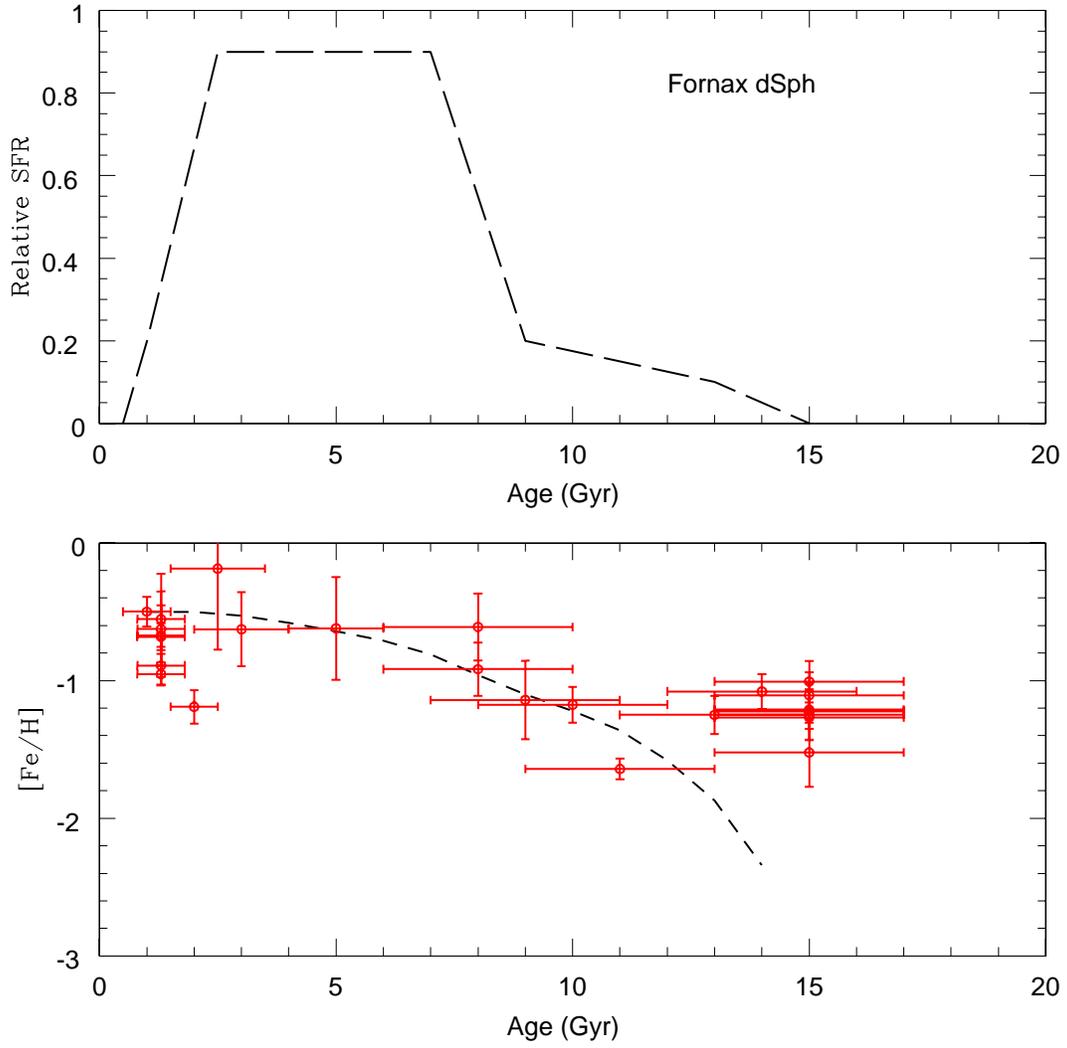,width=15cm}}}
\caption{
Here we display a possible star-formation and chemical evolution
scenario for Fornax over its entire history ($\sim$15~Gyr), as
described in Fig~14.
%In the
%upper panel is a schematic plot, consistent with all that we know of
%the stellar population of Fornax, of how the rate of star formation
%may have varied over time, back from the epoch of globular cluster
%formation around 15~Gyr ago. In the lower panel we plot a
%corresponding variation in metallicity over the same time frame.
%Overploted on the lower panel are our Ca~II triplet measurements for
%individual RGB stars, for which we determined ages using isochrones.
}
\label{fnxsfh}
\end{figure}

%Figure 18
\begin{figure}
\centerline{\hbox{\psfig{figure=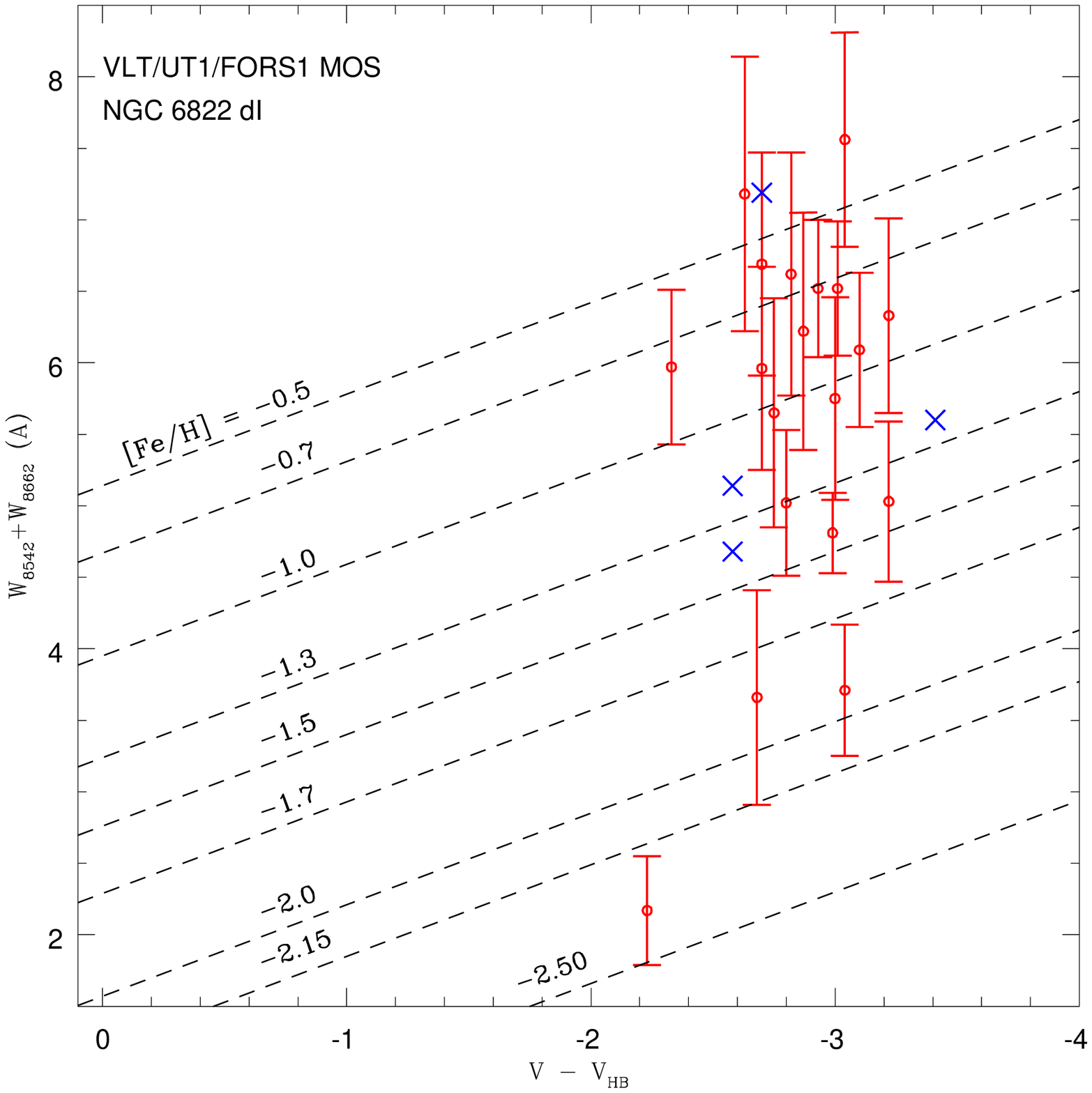,width=15cm}}}
\caption{
The summed equivalent width of the two stronger Ca~II triplet lines is
plotted, as described in Fig~12.
% against the V magnitude difference with the Horizontal Branch
%for the observed stars in the NGC~6822 dI.  Also plotted are lines of
%constant metallicity, as calibrated and checked with R97, in Figure~10
%(see \S 3.2 and \S 4.1).  The error bars come from the Gaussian
%fitting measurement errors.
}
\label{n6822res}
\end{figure}

\clearpage

%Figure 19
\begin{figure}
\centerline{\hbox{\psfig{figure=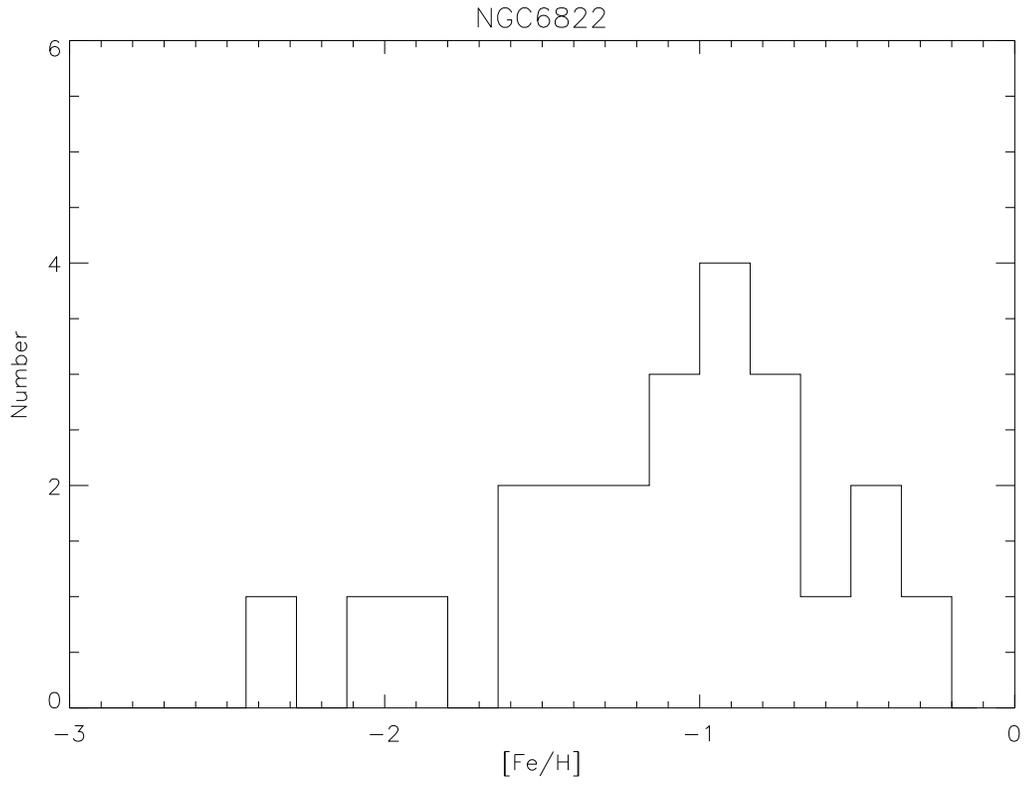,width=15cm}}}
\caption{
Here we plot the histogram distribution of NGC~6822 RGB Ca~II triplet
metallicities.
}
\label{n6822fehist}
\end{figure}

%Figure 20
\begin{figure}
\centerline{\hbox{\psfig{figure=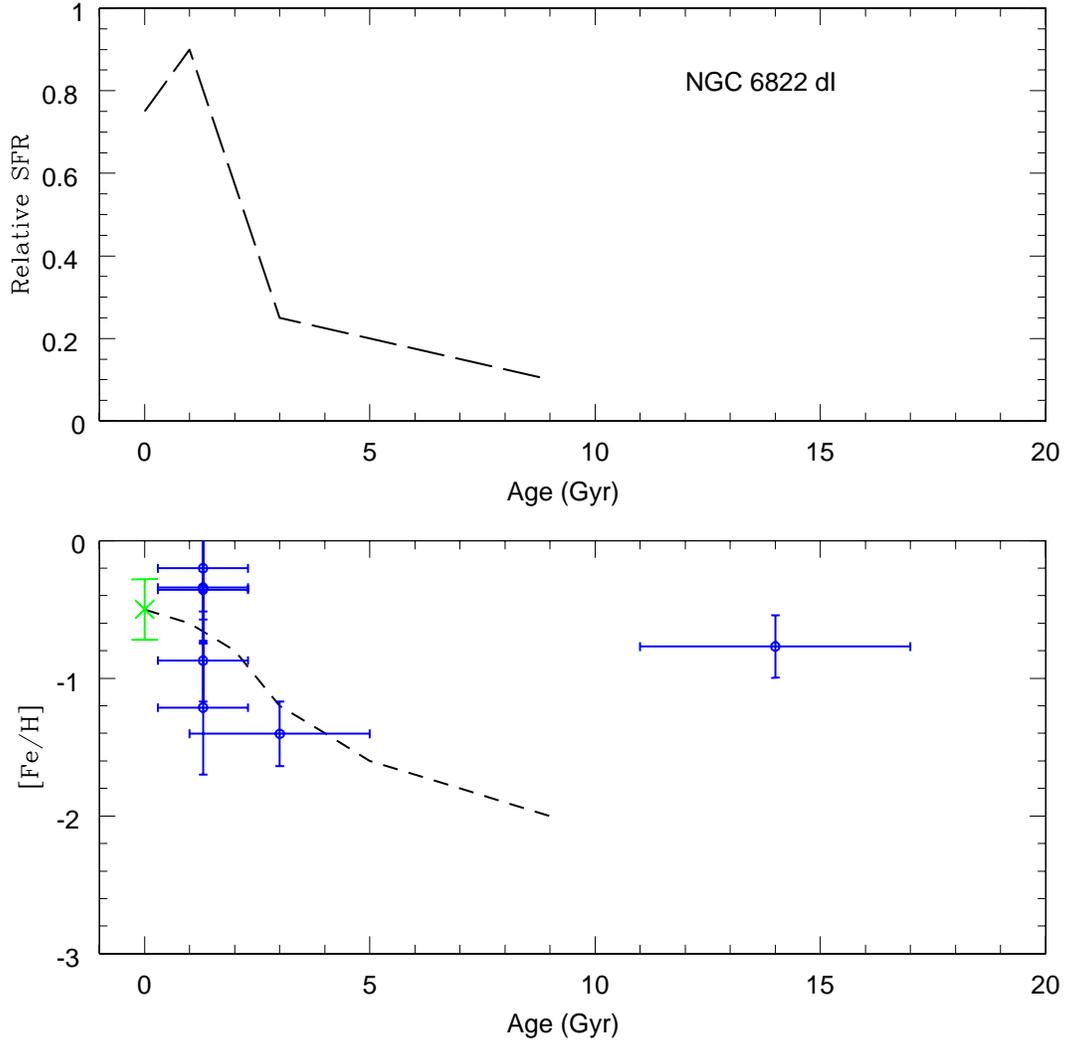,width=15cm}}}
\caption{
Here we display a highly speculative possible star-formation and
chemical evolution scenario for NGC~6822 over its entire history
($\sim$15~Gyr), as described in Fig~14. There is an additional
point for the present day metallicity measured from B-super-giant
spectra by Venn {\it et al.} (2001).
%In the upper panel is a schematic plot, consistent
%with all that we know of the stellar population of NGC~6822, of how
%the rate of star formation may have varied over time, back from the
%epoch of globular cluster formation around 15~Gyr ago. In the lower
%panel we plot a corresponding variation in metallicity over the same
%time frame.  Overploted on the lower panel are our Ca~II triplet
%measurements for individual RGB stars, for which we determined ages
%using isochrones.
}
\label{n6822sfh}
\end{figure}

\newpage
\clearpage

\centerline{Appendix: Tables giving positions of observed stars}

\centerline{\bf Table A1: Sculptor Positions}
\vskip-1cm

$$
\table
\tablespec{\l\l\l}
\body{
\skip{0.06cm}
\hline
\skip{0.025cm}
\hline
\skip{.2cm}
& Star & RA  & Dec (J2000) &\end
\skip{.1cm}
\hline
\skip{.05cm}
\hline
\skip{.45cm}
& c1-56 & 01:00:01.3 & $-$33:45:16 &\end 
& c1-70 & 01:00:02.9 & $-$33:44:59 &\end 
& c1-85 & 01:00:04.8 & $-$33:44:41 &\end 
& c1-68 & 01:00:02.9 & $-$33:44:06 &\end 
& c1-101& 01:00:06.9 & $-$33:43:45 &\end 
& c1-99 & 01:00:06.7 & $-$33:43:05 &\end 
& c1-43 & 00:59:59.7 & $-$33:42:37 &\end 
& c1-55 & 01:00:01.1 & $-$33:42:22 &\end 
& c1-67 & 01:00:02.8 & $-$33:42:00 &\end 
& c1-76 & 01:00:03.9 & $-$33:41:40 &\end 
& c1-81 & 01:00:04.6 & $-$33:41:13 &\end 
& c1-46 & 01:00:00.1 & $-$33:40:57 &\end 
& c1-47 & 01:00:00.1 & $-$33:40:41 &\end 
& c1-88 & 01:00:05.5 & $-$33:40:17 &\end 
& c1-78 & 01:00:04.0 & $-$33:40:02 &\end 
& c2-64 & 01:00:17.1 & $-$33:45:13 &\end 
& c2-81 & 01:00:19.0 & $-$33:44:52 &\end 
& c2-88 & 01:00:19.9 & $-$33:44:24 &\end 
& c2-73 & 01:00:18.0 & $-$33:44:12 &\end 
& c2-38 & 01:00:13.4 & $-$33:43:38 &\end 
& c2-39 & 01:00:13.5 & $-$33:43:20 &\end 
& c2-72 & 01:00:17.8 & $-$33:43:04 &\end 
& c2-52 & 01:00:15.5 & $-$33:42:37 &\end 
& c2-27 & 01:00:12.3 & $-$33:42:08 &\end 
& c2-60 & 01:00:16.8 & $-$33:41:46 &\end 
& c2-86 & 01:00:19.8 & $-$33:41:12 &\end 
& c2-82 & 01:00:19.5 & $-$33:40:36 &\end 
& c2-53 & 01:00:15.9 & $-$33:40:15 &\end 
& c2-85 & 01:00:19.8 & $-$33:40:04 &\end 
& o1-1  & 01:00:42.8 & $-$33:31:30 &\end 
& o1-4  & 01:00:43.3 & $-$33:30:37 &\end 
& o1-11 & 01:00:46.6 & $-$33:30:47 &\end 
& o1-14 & 01:00:48.0 & $-$33:30:38 &\end 
& o1-6  & 01:00:43.8 & $-$33:29:08 &\end 
& o1-15 & 01:00:49.1 & $-$33:29:45 &\end 
& o1-21 & 01:00:50.4 & $-$33:29:08 &\end 
& o1-22 & 01:00:50.8 & $-$33:27:52 &\end 
& o1-30 & 01:00:57.9 & $-$33:27:53 &\end 
& o2-28 & 00:59:29.0 & $-$33:49:12 &\end 
& o2-25 & 00:59:27.9 & $-$33:48:59 &\end 
& o2-33 & 00:59:31.8 & $-$33:47:45 &\end 
& o2-38 & 00:59:32.6 & $-$33:46:18 &\end 
& o2-46 & 00:59:33.8 & $-$33:45:05 &\end 
& o2-44 & 00:59:33.6 & $-$33:44:32 &\end 
\skip{.25cm}
\hline
\skip{0.025cm}
\hline
}
\endtable
$$

\newpage

\centerline{\bf Table A2: Fornax Positions}
\vskip-1cm

$$
\table
\tablespec{\l\l\l}
\body{
\skip{0.06cm}
\hline
\skip{0.025cm}
\hline
\skip{.2cm}
& Star & RA  & Dec (J2000) &\end
\skip{.1cm}
\hline
\skip{.05cm}
\hline
\skip{.45cm}
& c1-660 & 02:39:51.3 & $-$34:29:58 &\end
& c1-350 & 02:39:50.8 & $-$34:29:28 &\end
& c1-444 & 02:39:52.3 & $-$34:29:06 &\end
& c1-371 & 02:39:51.2 & $-$34:28:44 &\end
& c1-601 & 02:39:54.4 & $-$34:28:27 &\end
& c1-628 & 02:39:54.8 & $-$34:28:01 &\end
& c1-564 & 02:39:53.9 & $-$34:27:37 &\end
& c1-433 & 02:39:52.2 & $-$34:27:19 &\end
& c1-365 & 02:39:51.1 & $-$34:26:49 &\end
& c1-122 & 02:39:47.5 & $-$34:26:30 &\end
& c1-125 & 02:39:47.6 & $-$34:26:10 &\end
& c1-214 & 02:39:49.0 & $-$34:25:50 &\end
& c1-360 & 02:39:51.1 & $-$34:25:17 &\end
& c1-200 & 02:39:48.8 & $-$34:24:56 &\end
& c1-344 & 02:39:50.9 & $-$34:24:30 &\end
& c2-822 & 02:40:10.8 & $-$34:29:45 &\end
& c2-777 & 02:40:10.0 & $-$34:29:24 &\end
& c2-838 & 02:40:11.0 & $-$34:29:03 &\end
& c2-828 & 02:40:10.9 & $-$34:28:56 &\end
& c2-702 & 02:40:09.0 & $-$34:28:41 &\end
& c2-613 & 02:40:07.9 & $-$34:28:25 &\end
& c2-769 & 02:40:10.0 & $-$34:27:56 &\end
& c2-511 & 02:40:06.7 & $-$34:27:39 &\end
& c2-623 & 02:40:08.1 & $-$34:27:13 &\end
& c2-388 & 02:40:05.1 & $-$34:26:53 &\end
& c2-384 & 02:40:05.0 & $-$34:26:51 &\end
& c2-294 & 02:40:03.7 & $-$34:26:33 &\end
& c2-552 & 02:40:07.2 & $-$34:26:05 &\end
& c2-249 & 02:40:03.1 & $-$34:25:46 &\end
& c2-413 & 02:40:05.4 & $-$34:25:20 &\end
& c2-647 & 02:40:08.5 & $-$34:24:53 &\end
& c2-621 & 02:40:08.2 & $-$34:24:30 &\end
& c2-41	 & 02:40:00.2 & $-$34:24:29 &\end
\skip{.1cm}
\hline
\skip{.05cm}
\hline
}
\endtable
$$

\newpage

\centerline{\bf Table A3: NGC~6822 Positions}
\vskip-1cm

$$
\table
\tablespec{\l\l\l}
\body{
\skip{0.06cm}
\hline
\skip{0.025cm}
\hline
\skip{.2cm}
& Star & RA  & Dec (J2000) &\end
\skip{.1cm}
\hline
\skip{.05cm}
\hline
\skip{.45cm}
& s1-309 & 19:44:47.5 & $-$14:47:00 &\end
& s1-186 & 19:44:47.7 & $-$14:46:41 &\end
& s1-212 & 19:44:48.2 & $-$14:45:09 &\end
& s1-210 & 19:44:48.2 & $-$14:45:06 &\end
& s1-200 & 19:44:47.9 & $-$14:44:43 &\end
& s1-59  & 19:44:44.8 & $-$14:44:18 &\end
& s1-188 & 19:44:47.7 & $-$14:43:39 &\end
& s1-153 & 19:44:47.1 & $-$14:43:17 &\end
& s1-43  & 19:44:44.5 & $-$14:42:58 &\end
& s1-205 & 19:44:48.1 & $-$14:42:14 &\end
& s1-204 & 19:44:48.1 & $-$14:42:02 &\end
& s1-378 & 19:44:52.0 & $-$14:41:43 &\end
& s1-111 & 19:44:46.7 & $-$14:41:21 &\end
& s1b-280& 19:44:49.7 & $-$14:44:49 &\end
& s2-208 & 19:45:02.3 & $-$14:45:55 &\end
& s2-246 & 19:45:03.4 & $-$14:45:31 &\end
& s2-263 & 19:45:09.8 & $-$14:45:10 &\end
& s2-352 & 19:45:05.7 & $-$14:44:51 &\end
& s2-354 & 19:45:05.7 & $-$14:44:45 &\end
& s2-250 & 19:45:03.5 & $-$14:44:02 &\end
& s2-271 & 19:45:04.0 & $-$14:43:45 &\end
& s2-142 & 19:45:00.8 & $-$14:43:12 &\end
& s2-248 & 19:45:03.5 & $-$14:43:00 &\end
& s2-117 & 19:45:00.2 & $-$14:42:07 &\end
& s2-199 & 19:45:02.1 & $-$14:41:51 &\end
& s2-198 & 19:45:02.1 & $-$14:41:39 &\end
\skip{.1cm}
\hline	  
\skip{.05cm}
\hline	  
}	  
\endtable 
$$	  
	  
\end{document}